\documentstyle[12pt,aaspp4,epsfig,subfig,color]{aastex}
\def\HI{\hbox{H~$\scriptstyle\rm I\ $}}
\def\arcs{$''$}

\def\sfrd{\,{\rm M_\odot\,yr^{-1}\,Mpc^{-3}}}
\def\emunits{\,{\rm ergs\,s^{-1}\,Hz^{-1}\,Mpc^{-3}}}

\begin{document}

\title{Star Formation at z$\sim$6: $i$-dropouts in the ACS GTO
fields\altaffilmark{1,2}}

\author{R.J. Bouwens\altaffilmark{3,4},
        G.D. Illingworth\altaffilmark{3,5},
P. Rosati\altaffilmark{6},
C. Lidman\altaffilmark{6},
T. Broadhurst\altaffilmark{7},
M. Franx\altaffilmark{8},
H.C. Ford\altaffilmark{9},
D. Magee\altaffilmark{3},
N. Ben\'{\i}tez\altaffilmark{9},
J.P. Blakeslee\altaffilmark{9},
G.R. Meurer\altaffilmark{9},
M. Clampin\altaffilmark{10},
G.F. Hartig\altaffilmark{9},
D.R. Ardila\altaffilmark{9},
F. Bartko\altaffilmark{11}, 
R.A. Brown\altaffilmark{10},
C.J. Burrows\altaffilmark{10},
E.S. Cheng\altaffilmark{12},
N.J.G. Cross\altaffilmark{9},
P.D. Feldman\altaffilmark{9},
D.A. Golimowski\altaffilmark{9},
C. Gronwall\altaffilmark{13},
L. Infante\altaffilmark{14},
R.A. Kimble\altaffilmark{12},
J.E. Krist\altaffilmark{9},
M.P. Lesser\altaffilmark{15},
A.R. Martel\altaffilmark{9},
F. Menanteau\altaffilmark{9},
G.K. Miley\altaffilmark{8},
M. Postman\altaffilmark{9},
M. Sirianni\altaffilmark{9}, 
W.B. Sparks\altaffilmark{2}, 
H.D. Tran\altaffilmark{9}, 
Z.I. Tsvetanov\altaffilmark{9},   
R.L. White\altaffilmark{9,10}
\& W. Zheng\altaffilmark{9}}

\altaffiltext{1}{Based on observations made with the NASA/ESA Hubble
Space Telescope, which is operated by the Association of Universities
for Research in Astronomy, Inc., under NASA contract NAS
5-26555. These observations are associated with programs
\#7817,9290,9301,9583.}

\altaffiltext{2}{Based on observations collected at the European
 Southern Observatory, Paranal, Chile (LP166.A-0701)}

\altaffiltext{3}{Astronomy Department,
    University of California,
    Santa Cruz, CA 95064}

\altaffiltext{4}{bouwens@ucolick.org}
\altaffiltext{5}{gdi@ucolick.org}



\altaffiltext{6}{European Southern Observatory,
Karl-Schwarzschild-Strasse 2, D-85748 Garching, Germany.}

\altaffiltext{7}{Racah Institute of Physics, The Hebrew University,
Jerusalem, Israel 91904.}


\altaffiltext{8}{Leiden Observatory, Postbus 9513, 2300 RA Leiden,
Netherlands.}


\altaffiltext{9}{Department of Physics and Astronomy, Johns Hopkins
University, 3400 North Charles Street, Baltimore, MD 21218.}

\altaffiltext{10}{STScI, 3700 San Martin Drive, Baltimore, MD 21218.}


\altaffiltext{11}{Bartko Science \& Technology, P.O. Box 670, Mead, CO
80542-0670.}


\altaffiltext{12}{NASA Goddard Space Flight Center, Laboratory for
Astronomy and Solar Physics, Greenbelt, MD 20771.}


\altaffiltext{13}{Department of Astronomy and Astrophysics, The
Pennsylvania State University, 525 Davey Lab, University Park, PA
16802.}


\altaffiltext{14}{Departmento de Astronom\'{\i}a y Astrof\'{\i}sica,
Pontificia Universidad Cat\'{\o}lica de Chile, Casilla 306, Santiago
22, Chile.}


\altaffiltext{15}{Steward Observatory, University of Arizona, Tucson,
AZ 85721.}

\begin{abstract}

Using an $i-z$ dropout criterion, we determine the space density of
$z\sim6$ galaxies from two deep ACS GTO fields with deep optical-IR
imaging.  A total of 23 objects are found over 46 arcmin$^2$, or
$\sim$0.5$\pm$0.1 objects arcmin$^{-2}$ down to $z_{AB}\sim27.3$
(6$\sigma$), or a completeness-corrected $\sim$0.5$\pm$0.2 objects
arcmin$^{-2}$ down to $z_{AB}\sim26.5$ (including one probable
$z\sim6$ AGN).  Combining deep ISAAC data for our RDCS1252-2927 field
($J_{AB}\sim25.7$ and $Ks_{AB}\sim25.0$ (5 $\sigma$)) and NICMOS data
for the HDF North ($J_{110,AB}$ and $H_{160,AB}$ $\sim$ 27.3 (5
$\sigma$)), we verify that these dropouts have relatively flat
spectral slopes as one would expect for star-forming objects at
$z\sim6$.  Compared to the average-color ($\beta=-1.3$) $U$-dropout in
the Steidel et al.\ (1999) $z\sim3$ sample, $i$-dropouts in our sample
range in luminosity from $\sim$1.5 $L_{*}$ ($z_{AB}\sim25.6$) to
$\sim$0.3 $L_{*}$ ($z_{AB}\sim27.3$) with the exception of one very
bright candidate at $z_{850,AB}\sim24.2$.  The half-light radii vary
from 0.09\arcs$~$to 0.29\arcs, or 0.5 kpc to 1.7 kpc.  We derive the
$z\sim6$ rest-frame UV luminosity density (or star formation rate
density) using three different procedures.  All three procedures make
use of simulations based on a slightly lower redshift ($z\sim5$)
$V_{606}$-dropout sample from CDF South ACS images.  First, we make a
direct comparison of our findings with a no-evolution projection of
this $V$-dropout sample, allowing us to automatically correct for the
light lost at faint magnitudes or lower surface brightnesses.  We find
23$\pm$25\% more $i$-dropouts than we predict, consistent with no
strong evolution over this redshift range.  Adopting previous results
to $z\sim5$ (Bouwens, Broadhurst, \& Illingworth 2003; Thompson et
al.\ 2001), this works out to a mere $20\pm29$\% drop in the
luminosity density from $z\sim3$ to $z\sim6$.  Second, we use the same
$V$-dropout simulations to derive a detailed selection function for
our $i$-dropout sample and compute the $UV$-luminosity density
($7.2\pm2.5\times10^{25}\emunits$ down to $z_{AB}\sim27$).  We find a
$39\pm21$\% drop over the same redshift range ($z\sim3-6$), consistent
with the first estimate.  This is our preferred value and suggests a
star formation rate of $0.0090\pm0.0031\,\sfrd$ to $z_{AB}\sim27$, or
$\sim0.036\pm0.012\,\sfrd$ extrapolating the luminosity function to
the faint limit.  Third, we follow a very similar procedure, except
that we assume no incompleteness, finding a rest-frame continuum
luminosity density which is $\sim2-3\times$ lower than our other two
determinations.  This final estimate is to be taken as a lower limit,
and is important in the event that there are modest changes in the
colors or surface brightnesses from $z\sim5$ to $z\sim6$ (the other
estimates assume no large changes in the intrinsic selectability of
objects).  We note that all three estimates are well within the
canonical range of luminosity densities (e.g., Madau, Haardt, \& Rees
1999) necessary for reionization of the universe at this epoch by
star-forming galaxies.

\end{abstract}


\section{Introduction}

The Hubble Deep Field (HDF) campaign has been highly influential in
shaping our understanding of star formation in the high-redshift
universe (Williams et al.\ 1996; Casertano et al.\ 2000; Ferguson,
Dickinson, \& Williams 2000).  Early results demonstrated that the
star formation rate density--as measured from the rest-frame continuum
UV--increased from $z\sim4$ to an apparent peak around $z\sim1-3$
(Lilly et al.\ 1996; Madau et al.\ 1996; Connolly et al.\ 1997; Cowie
et al.\ 1999).  While these results were largely solidified by Steidel
et al.\ (1999) with his wide-area $U$-dropout survey, the addition of
NICMOS data to the HDF North demonstrated that this trend continued to
$z\sim6$ (Thompson et al.\ 1999; Deltorn et al.\ 2003; Bouwens,
Broadhurst, \& Illingworth 2003, hereafter denoted BBI).

Unfortunately, these studies were limited enough in area to raise
doubts about how representative they really were of the high-redshift
universe.  They also suffered from a lack of deep high-resolution
imaging at wavelengths intermediate between the optical regime and the
near infrared, necessary for obtaining a more detailed look at the
$z\sim6-7$ universe.  Deep NICMOS observations have been useful in
addressing this latter shortcoming, but only partially due to its
small field of view.  Fortunately, the installation of the Advanced
Camera for Surveys (ACS) (Ford et al.\ 1998) on the Hubble Space
Telescope has helped to redress several of these issues, including
crucially for the first time imaging in the $z$-band, permitting a
more secure detection of objects at high redshift ($z\sim5-6$).
Moreover, its 10$\times$ improvement over WFPC2 in surveying
capability allows large areas to be surveyed to nearly HDF depths
(Ford et al.\ 2003), the Great Observatories Origins Deep Survey
(GOODS) being a notable example (Dickinson \& Giavalisco 2002).

Here, we describe some early work done using deep ACS data to extend
these searches to $z\sim6$, to establish the prevalence of galaxies in
this era.  Interest in star formation at $z\sim6$ has been
particularly intense as of late because of recent absorption line
studies on 3 QSOs at $z>5.8$, suggesting that reionization may have
happened at about this epoch (Fan et al.\ 2001; Becker et al.\ 2001;
Fan et al.\ 2002).  In this work, we consider two fields from the ACS
GTO program, RDCS1252-2927 and the HDF North, in our search for
$z\sim6$ objects.  Both fields have deep ACS $i$ and $z$ data, and
infrared observations, important for securely identifying $z\sim6$
objects.  Relative to other work (Yan et al.\ 2003; Stanway, Bunker,
\& McMahon 2003), the present search is slightly deeper, with better
IR data to confirm the redshift identifications.  In fact, our use of
the HDF North field is especially propitious, given the exceptionally
deep WFPC2 and NICMOS images available to examine faint $z\sim6$
candidates.  We put this new population in context by comparing them
with lower-redshift expectations, projecting $z\sim5$ galaxy samples
from CDF South GOODS to $z\sim6$ using our cloning formalism
previously used in work on the HDFs (Bouwens, Broadhurst, \& Silk
1998a,b; BBI).

We begin by presenting our data sets, describing our procedure for
doing object detection and photometry, and finally discussing our
$z\sim6$ $i$-dropout selection criterion (\S2).  In \S3, we present
our results.  In \S4, we describe a comparison against the wide-area
GOODS sample and then use these simulations to make three different
estimates of the $z\sim6$ rest-frame continuum $UV$ luminosity density
(\S5).  Finally, in $\S6$ and $\S7$, we discuss and summarize our
findings.  Note that we denote the $F775W$, $F850LP$, $F110W$, and
$F160W$ bands as $i_{775}$, $z_{850}$, $J_{110}$, and $H_{160}$,
respectively, and we assume [$\Omega_M$,$\Omega_{\Lambda}$,$h$] =
[0.3,0.7,0.7] in accordance with the recent Wilkinson Microwave
Anisotropy Probe (WMAP) results (Bennett et al.\ 2003).

\section{Observations}

\subsection{Data}

There are two different fields from our ACS GTO program which are
particularly useful for $i$-dropout searches.  The first involves deep
ACS WFC $i_{775}$ and $z_{850}$ images of RDCS1252-2927, a $z=1.235$
cluster.  RDCS1252-2927 was selected from the ROSAT Deep Cluster
Survey (Rosati et al.\ 1998; Rosati et al.\ 2003).  Three orbits in
$i_{775}$ and five orbits in $z_{850}$ were obtained at four
overlapping pointings, arranged in a 2x2 grid with an overlap of
$\sim$1 arcmin so that the overlapping regions ($\sim$10 arcmin$^2$)
were covered to a depth of six orbits in $i_{775}$/ten orbits in
$z_{850}$ with a small central region ($\sim$1 arcmin$^2$) being
covered to a depth of twelve orbits in $i_{775}$/twenty orbits in
$z_{850}$.  The ACS images were aligned, cosmic-ray rejected, and
drizzled together using the ACS GTO pipeline (Blakeslee et al.\
2003a).

Very deep integrations were obtained on ISAAC over 4 overlapping
regions on RDCS1252-2927 (covering 4x4 arcmin, or $\sim$44\% of our 36
arcmin$^2$ ACS mosaic).  A total of 24.1 and 22.7 hours were invested
in the $J$ and $Ks$ integrations, respectively ($\sim$6 hrs and
$\sim$5.8 hrs at each of the 4 offset positions).  These observations
reached $J_{AB} = 25.7$ ($5\sigma$) and $Ks_{AB} = 25.0$ ($5\sigma$)
in the shallower, non-overlapping regions and $J_{AB} = 26.5$
($5\sigma$) and $Ks_{AB} = 25.8$ ($5\sigma$) in the small ($1' \times
1'$) central region, with a FWHM for the PSF which was almost
uniformly $\sim0.45$\arcs$~$ across the entire IR mosaic.  These data
were then aligned with our optical data and resampled onto the same
0.05\arcs-pixel grid.

The second field utilizes deep ACS observations of the HDF North,
taken as part of our GTO program, 2.5 orbits in $i_{775}$ and 4.5
orbits in $z_{850}$.  This data is supplemented with 1.5 orbits in
$i_{775}$ and three orbits in $z_{850}$ from the GOODS program in this
field (representing three epochs of the GOODS program) to yield a
total depth of 4 orbits in $i_{775}$ and 7.5 orbits in $z_{850}$ over
an effective area of 10 arcmin$^2$.  To complement the ACS $i$ and $z$
data, both the HDF North optical data (Williams et al.\ 1996) and $JH$
infrared data from the Dickinson (1999) campaign were aligned and
registered onto the same 0.05\arcs-pixel scale as our ACS fields,
leaving the WFPC2 data with a FWHM of $\sim0.18$\arcs$~$ for the PSF
and the NICMOS data with a FWHM of $\sim0.25$\arcs$~$.  The NICMOS
images reached $J_{110,AB}\sim27.3$ ($5\sigma$) and
$H_{160,AB}\sim27.3$ (5$\sigma$).

The $i_{775,AB}=25.64$ and $z_{850,AB}=24.84$ CALACS (02/20/03)
zeropoints (Siranni et al.\ 2003) are assumed throughout, along with a
galactic absorption of $E(B-V)$=0.075 and 0.012 for the two fields
(from the Schlegel, Finkbeiner, \& Davis 1998 extinction maps),
resulting in a correction of -0.11$^m$ and -0.02$^m$ to the $z_{850}$
zeropoint for RDCS1252-2927 and the HDF North, respectively (and a
correction of -0.15$^m$ and -0.024$^m$ for the $i_{775}$ filter).

\subsection{Detection and Photometry}

Briefly, object detection is performed on the basis of our deep WFC
$z_{850}$ images after smoothing the images with a 0.09\arcs-FWHM
Gaussian kernel and looking for $4.5 \sigma$ peaks.  Photometry is
obtained for all detected objects with SExtractor (Bertin \& Arnouts
1996) using two scalable apertures, the inner one to measure colors
and the outer aperture to estimate the total flux.  For both sets of
IR data, similarly-scaled apertures are used to measure colors, but a
correction is applied based on the $z_{850}$ image to estimate how
much flux is lost in the IR due to PSF smoothing.  Due to correlation
in the noise, a concerted attempt was made to model the noise so that
a relatively realistic treatment of uncertainties could be applied
throughout the analysis (Appendix A).  A more comprehensive
description of our techniques for object selection and photometry is
given in BBI.\footnote{Note that our procedure for object detection
and photometry is different from that used by the GTO team (Blakeslee
et al.\ 2003a; Ben{\' i}tez et al.\ 2003).}

In total, 4632 and 1261 objects were recovered in the RDCS1252-2927
and HDF North fields, respectively.  Most spurious detections were
eliminated by demanding that each object be a 6$\sigma$ detection
within one Kron-radius (Kron 1980) (typically $\sim0.15$\arcs).  Areas
contaminated by optical ghosts or satellite trails were not included
in the analysis (the excluded area was $<0.5$\%).  A number of
spurious detections were found around bright stars or ellipticals, a
problem exacerbated by the rather extended wings on the $z_{850}$ PSF.
After some preliminary cleaning, all point-sources were removed from
our catalogs ($\sim503$ and $\sim110$), the rationale being to
eliminate very red stars which might otherwise masquerade as
high-redshift objects.  We found that the SExtractor stellarity
parameter adequately identified stellar objects.  While such a cut
might eliminate genuine $z\sim6$ star-forming objects, all of the very
red ($(i_{775}-z_{850})_{AB}>1.5$) point-like objects ($\sim4$) were
found to have much redder $z-J$ colors ($\sim1.2$) than most of the
probable $z\sim6$ objects, and therefore a stellar identification
seemed reasonable (see the systematic color differences between
point-like and extended $(i_{775}-z_{850})_{AB}>1.5$ objects presented
in Figure 1).  We note that an examination of red $i-z>1.3$ point-like
objects in our fields revealed one probable $z\sim6$ AGN (Appendix B).

\subsection{Dropout Selection}

The Lyman-break technique takes advantage of the increasingly strong
deficit of flux at high redshift caused by the intervening Lyman-alpha
forest eating into the spectrum shortward of 1216 $\textrm{\AA}$
(Madau 1995).  Combining this with flux information redward of the
break permits one to determine the spectral slope redward of the break
and therefore relatively robustly distinguish the objects of interest
from intrinsically red galaxies at lower redshift, as demonstrated by
extensive spectroscopic work done on a variety of different dropout
samples (Steidel et al.\ 1996a,b; Steidel et al.\ 1999; Weymann et
al.\ 1998; Fan et al.\ 2001).

For our filter set, the $i_{775}-z_{850}$ color measures the spectral
break and the $z_{850}-J$ color defines the spectral slope redward of
this break.  In Figures 1-2, we illustrate how a starburst spectrum
($100$ Myr continuous star formation) attenuated with various
opacities of dust ($E(B-V)=0,0.2,0.4$) would move through this
color-color space as a function of redshift.  In both plots, it is
clear that beyond $z\sim5.5$, the template $i-z$ colors become very
red ($>1.2$) while the $z-J$ colors remain very blue ($<0.5-1$).  For
reference, we also include the colors of possible lower redshift
interlopers using the Coleman, Wu, \& Weedman (1976) spectral
templates.

\begin{figure}
\epsscale{0.95}
\plotone{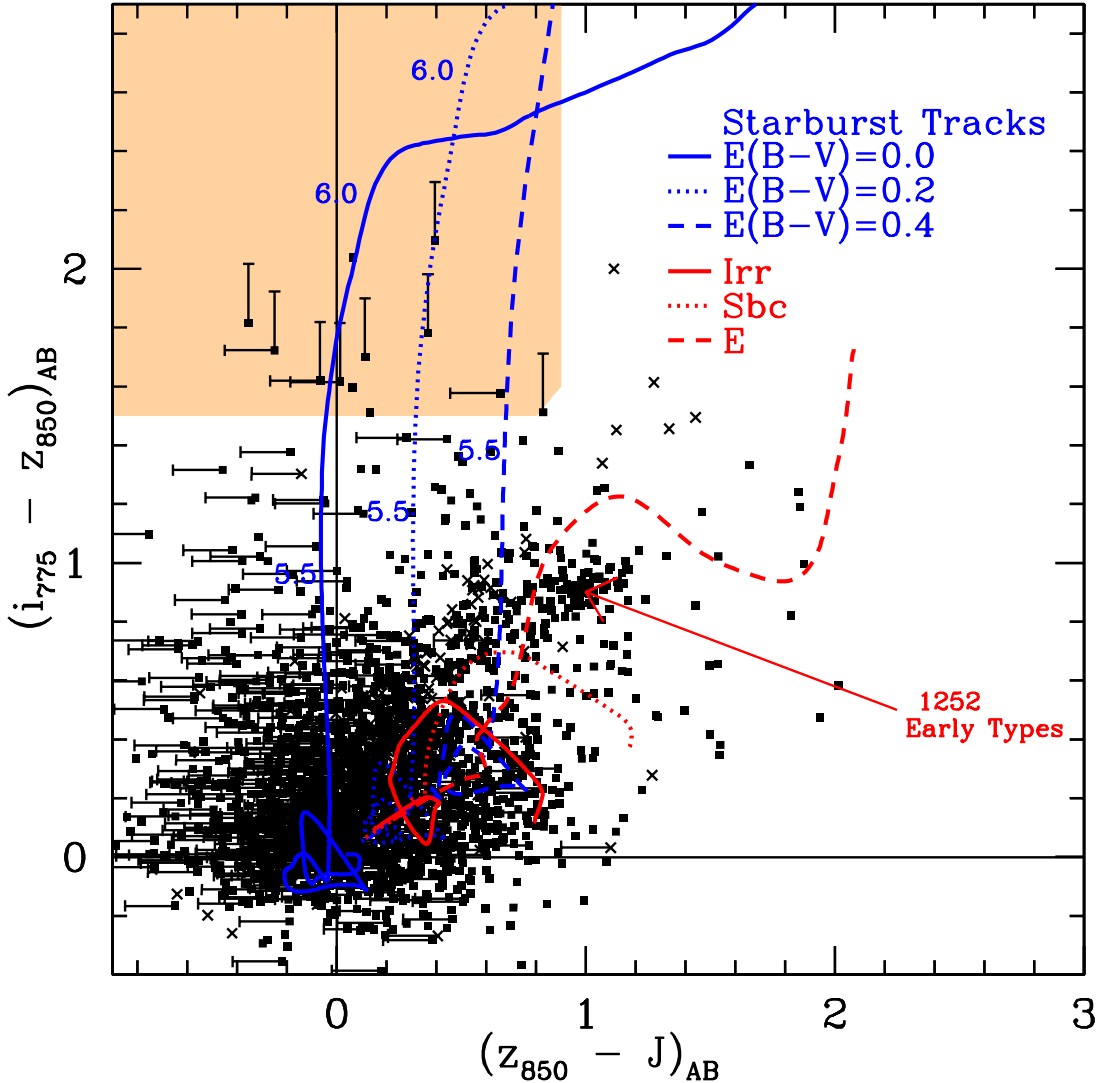}
\caption{$(i_{775}-z_{850})_{AB}$ vs. $(z_{850} - J)_{AB}$
colour-colour diagram illustrating the position of our RDCS1252-2927
$i$-dropout sample (shaded region) relative to the photometric sample
as a whole.  Tracks for a $10^8$ year starburst with various amounts
of extinctions have been included to illustrate both the typical
redshifts (labelled for $z=5.5$ and $z=6$) and SED types included in
the selection window.  The low-redshift ($0<z<1.2$) tracks for typical
E, Sbc, and Irr spectra have been included as well to illustrate the
region in colour-colour space where possible contaminants might lie.
There is a clear separation between the $i-z>1.3$ point-like objects
(crosses) and $i-z>1.3$ extended objects (solid squares) along the
$z-J$ axis.  The distribution of objects in color-color space led us
to adopt $(i-z)>1.5$ as our generalized $i$-dropout selection
criteria.  In all cases, error bars represent 2 $\sigma$ limits.  The
clump at $(i_{775}-z_{850})_{AB}\sim0.9$ and $(z_{850}-J)_{AB}\sim1$
are early-type galaxies from the cluster.}
\end{figure}

\begin{figure}
\epsscale{0.95}
\plotone{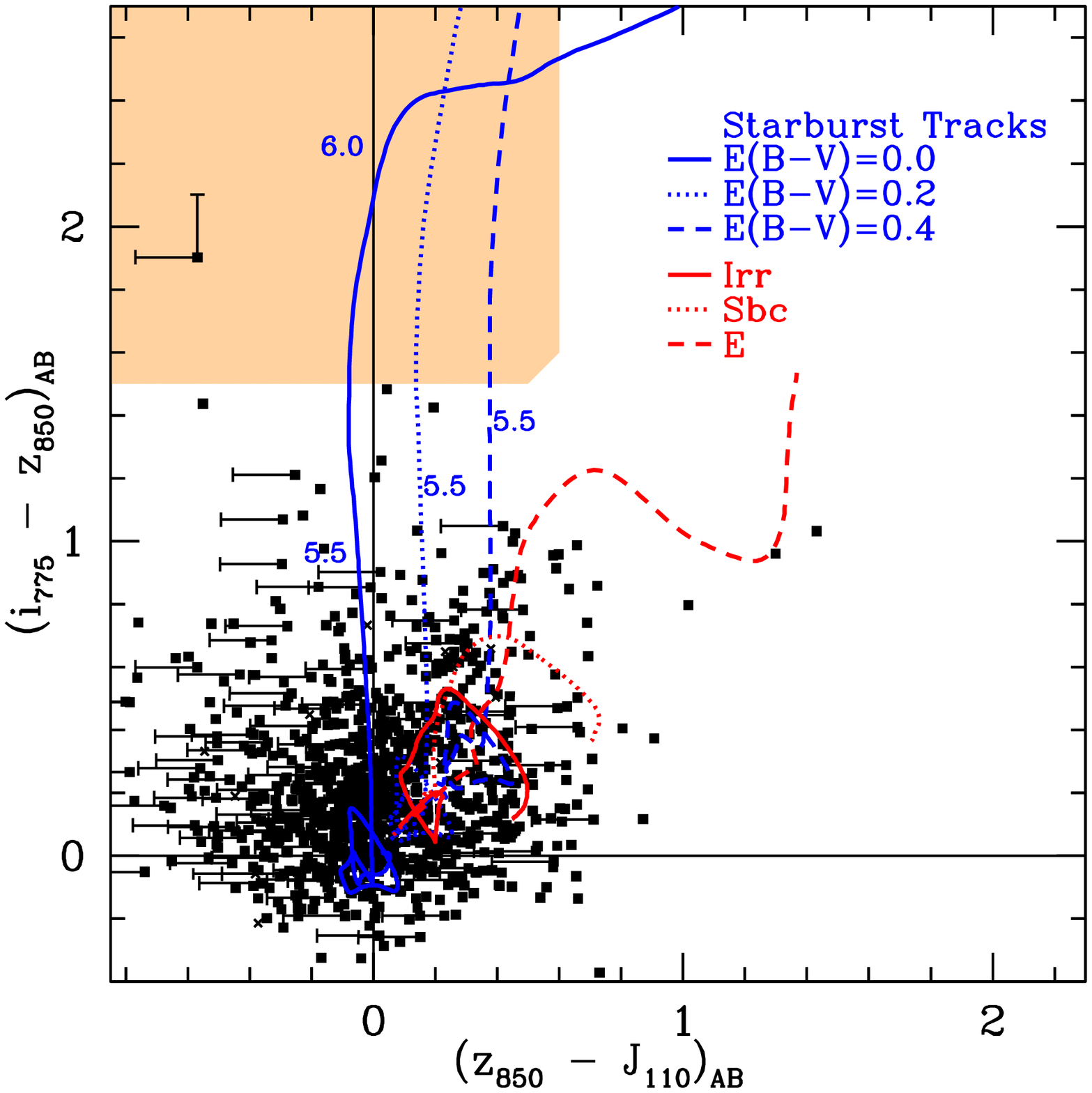}
\caption{Same as Figure 1, but for $i$-dropouts from the HDF North.
Note that the NICMOS $J_{110,AB}$ filter shown here is distinct from
the ground-based $J_{AB}$ filter used in Figure 1.}
\end{figure}

After considerable experimentation, we adopted a simple
$(i_{775}-z_{850})_{AB}>1.5$ cut throughout in our selection of
$i$-dropouts.  This color cut is motivated by evaluating object
selection in regions where we have both ACS and infrared data.  We
present such data in Figures 1-2 from our RDCS1252-2927 and HDF North
fields, providing 2$\sigma$ upper-limits for objects without
significant infrared flux.  We have also lightly shaded those regions
in $i-z$, $z-J$ color space where $z>5.5$ $i$-dropouts are expected to
lie: $(i_{775} - z_{850})_{AB} > 1.5$, $(i_{775} - z_{850})_{AB} >
(z_{775} - J)_{AB} + 0.7$, $(z_{775} - J)_{AB} < 0.9$, $z_{850,AB} <
27.3$ for RDCS1252-2927 and $(i_{775} - z_{850})_{AB} > 1.5$,
$(i_{775} - z_{850})_{AB} > (z_{775} - J_{110})_{AB} + 1.0$, $(z_{775}
- J_{110})_{AB} < 0.6$, $z_{850,AB} < 27.3$ for the HDF North.  A
quick glance shows that objects with very red ($>1.5$) $i-z$ colors
also have blue $z-J$ colors and lie exclusively in this region,
thereby validating our basic selection criteria.  Over the 21
arcmin$^2$ where we have infrared coverage, we find 11 objects in
RDCS1252-2927 and 1 object in the HDF North which satisfy our
$i-z>1.5$ cut.  $i_{775}z_{850}JKs$ and $i_{775}z_{850}J_{110}H_{160}$
images for these objects are shown in Figure 3, along with plots
showing their position in color-color space, fits of plausible SEDs to
the broadband fluxes, and an estimated redshift.  The photometric
redshifts are estimated using a bayesian formalism similar to that
outlined in Ben{\' i}tez (2000), using a prior which matches the
observed distribution of $z\sim3$ spectral slopes (Steidel et al.\
1999).  Our typical $i$-dropout has a S/N of $\sim2-3$ in the
infrared, more than adequate to put good constraints on the spectral
slope redward of the break.  To illustrate this and to indicate how
different the $i$-dropouts really are from possible low redshift
contaminants, we show three $i-z>1.3$ objects with red $z-J>0.8$
colors in Figure 4.  For reference, we also include a figure with the
$z=5.60$ Weymann et al.\ (1998) object from the HDF North to show the
position of a spectroscopically-confirmed $z>5.5$ object on these
diagrams (Figure 5).

Having presented our entire sample of $i$-dropouts with infrared
coverage, we now move onto quantifying the contamination rate due to
low redshift interlopers.  The most obvious way of doing this is
simply to count the fraction of objects with $i-z>1.5$ which satisfy
the two-color criteria we specified above versus those that do not.
Unfortunately, for many objects we only have limits and not precise
measures of the IR colors, leaving us with cases where we are not sure
if an object lies in our sample or not.  Therefore, following Pozzetti
et al.\ (1998) in their analysis of dropouts in the HDF North, we
resort to the use of Kaplan-Meier estimators with censoring (Lavalley,
Isobe, \& Feigelson 1992), where the implicit assumption is that
censoring is random, e.g., that objects with and without limits are
drawn from the same parent distribution.  Given the narrow range of
magnitudes and sizes in our sample we believe this assumption to be
approximately satisfied.  Performing this analysis on all objects in
our fields with $i-z>1.3$ (46 objects), we find the following
contamination fraction as a function of $(i-z)$ color: 0\%
$(i-z)>1.7$, 13\% $(i-z)>1.5$, and 21\% $(i-z)>1.3$.  (Note that this
estimate is for samples from which the point sources (stars or AGNs)
have already been removed.)  While our data set contains two
$i-z\sim1.3-1.4$ objects which are apparently ellipticals or Extremely
Red Objects (EROs) at $\sim$25.3, $z-K \sim 2$ and $z-K \sim 3$
(Figure 4), all the objects with $(i-z)>1.5$ have infrared colors
consistent with their being at high redshift.

This small contamination fraction (13\%) allows us to substantially
increase the size of our $z\sim6$ sample by including
$(i_{775}-z_{850})_{AB}>1.5$ objects without IR coverage.  A total of
11 such objects satisfy these criteria: 10 from RDCS1252-2927 and 1
from the HDF North.

\begin{subfigures}
\begin{figure}
\epsscale{0.9}
\plotone{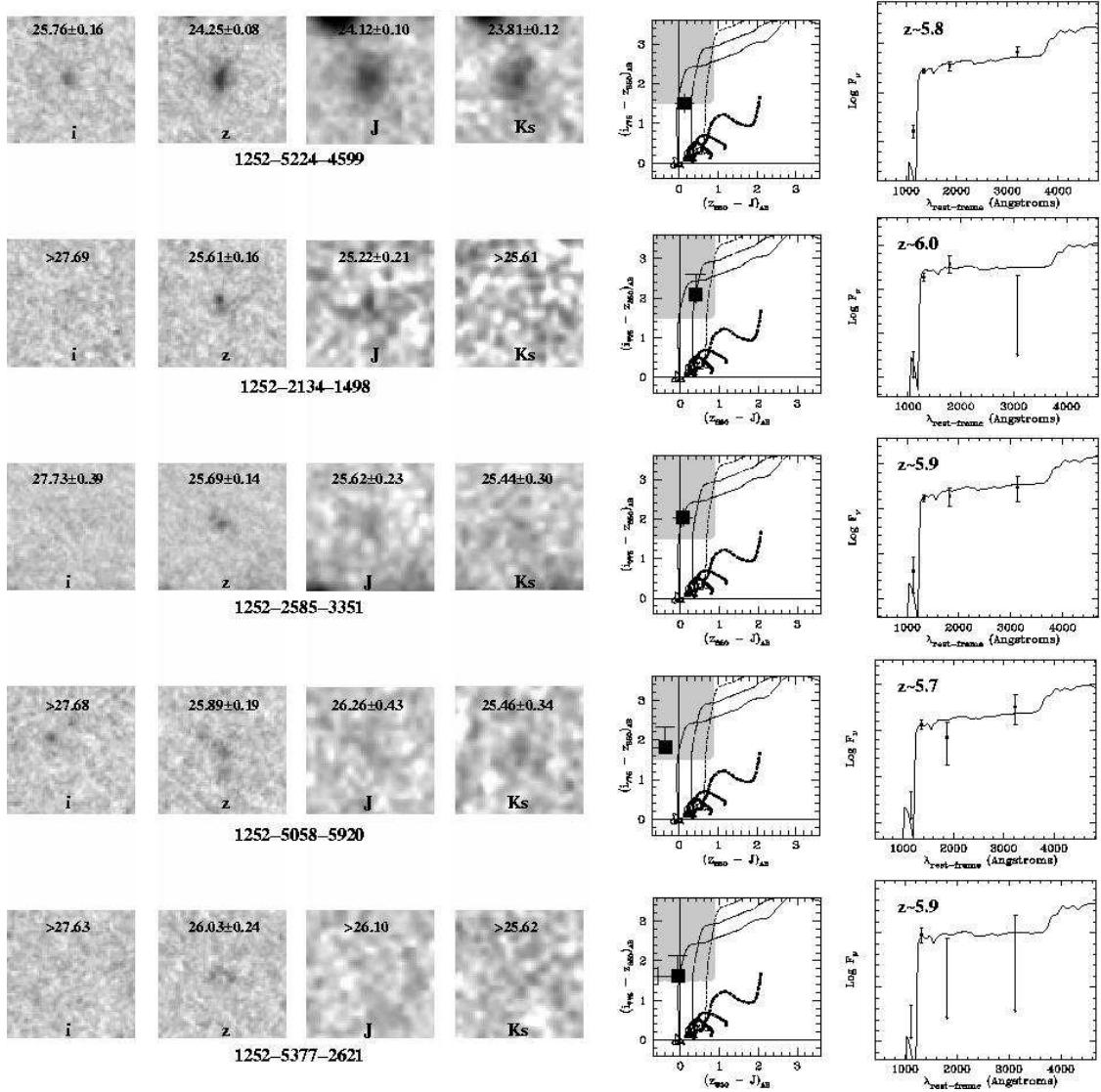}
\caption{$i_{775}z_{850}JK$ postage stamps of the 12 $i$-dropouts
($z\sim6$) identified in our ACS fields over the 21 arcmin$^2$ where
we had infrared coverage (11 $i$-dropouts did not have measurements in
the IR).  $i_{775}z_{850}J_{110}H_{160}$ postage stamps are shown for
the object from the HDF North.  The optical and IR images are smoothed
with 3$\times$3 and 6$\times$6 boxcars, respectively.  The position in
$i-z$, $z-J$ space is also indicated along with the broadband SEDs and
estimated redshift.  As in Figures 1-2, we have included lines
denoting the way starburst objects ($10^8$ yr bursts) with various
dust attentuations would move through color-color space as a function
of redshift.  We have also included the tracks of possible
interlopers.  As in Table 1, the ``1252-'' prefix denotes an object
from RDCS1252-2927.  The postage stamps are 
3.0\arcs$\times$3.0\arcs$\,\,$in size.}
\end{figure}

\begin{figure}
\epsscale{0.9}
\plotone{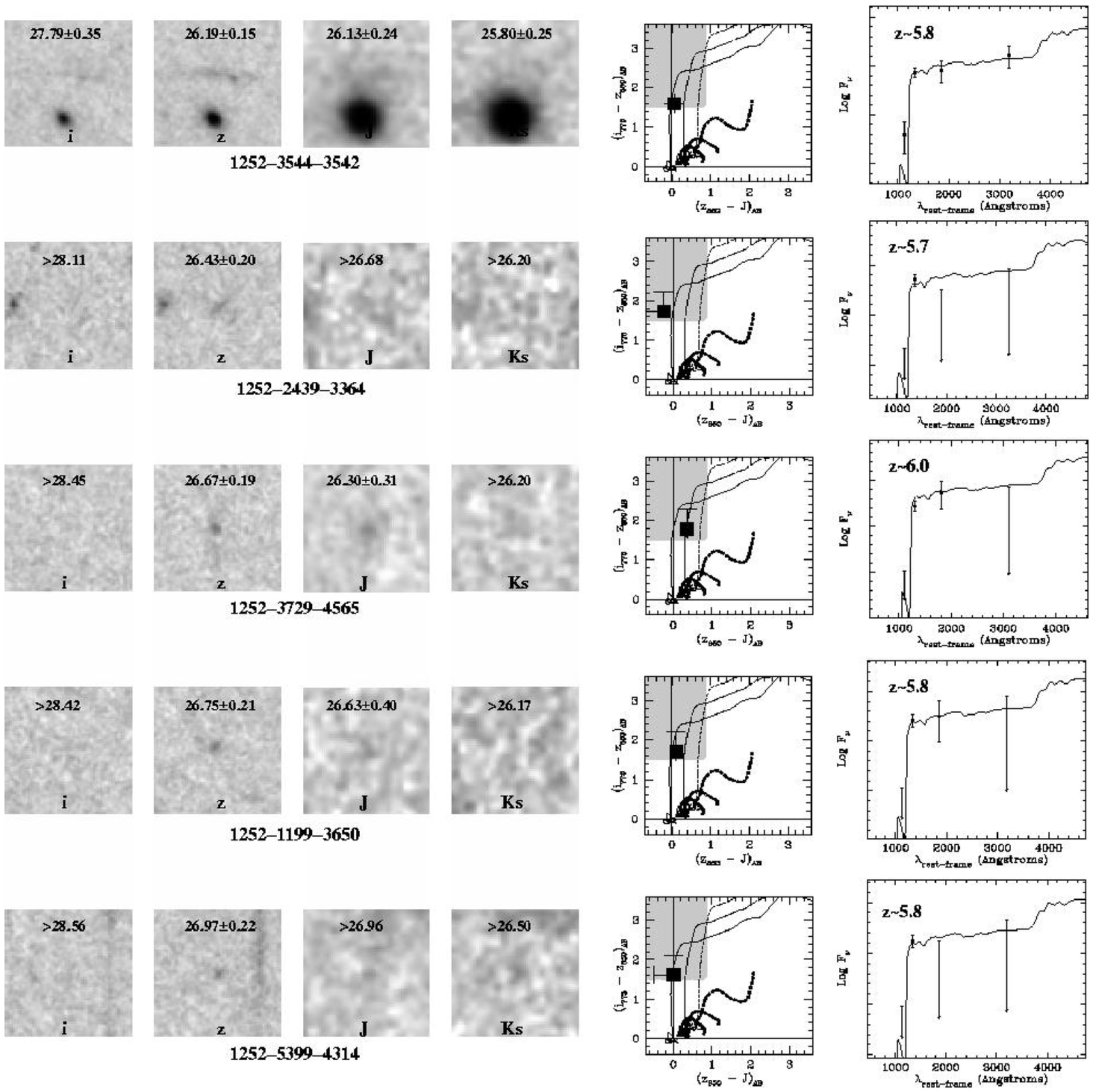}
\caption{}
\end{figure}

\begin{figure}[p]
\epsscale{0.9}
\plotone{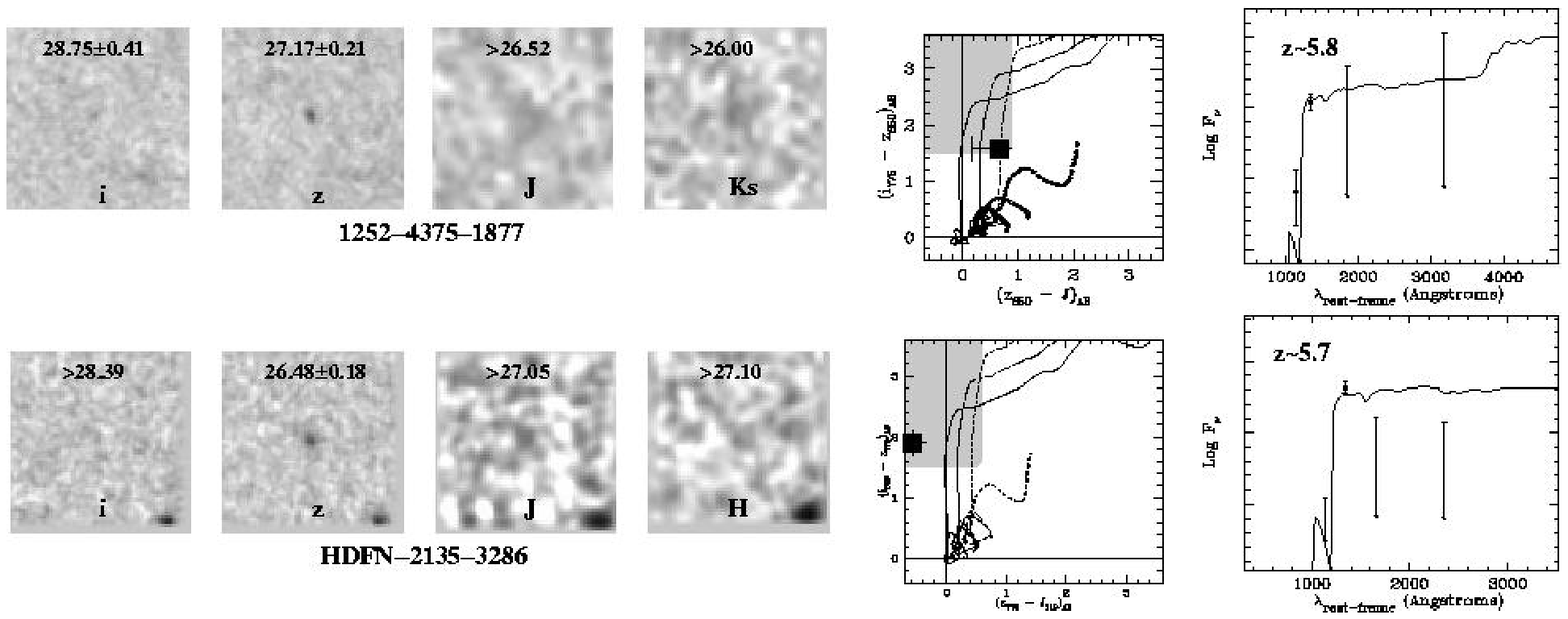}
\caption{}
\end{figure}
\end{subfigures}

\clearpage
\begin{figure}[p]
\epsscale{0.9}
\plotone{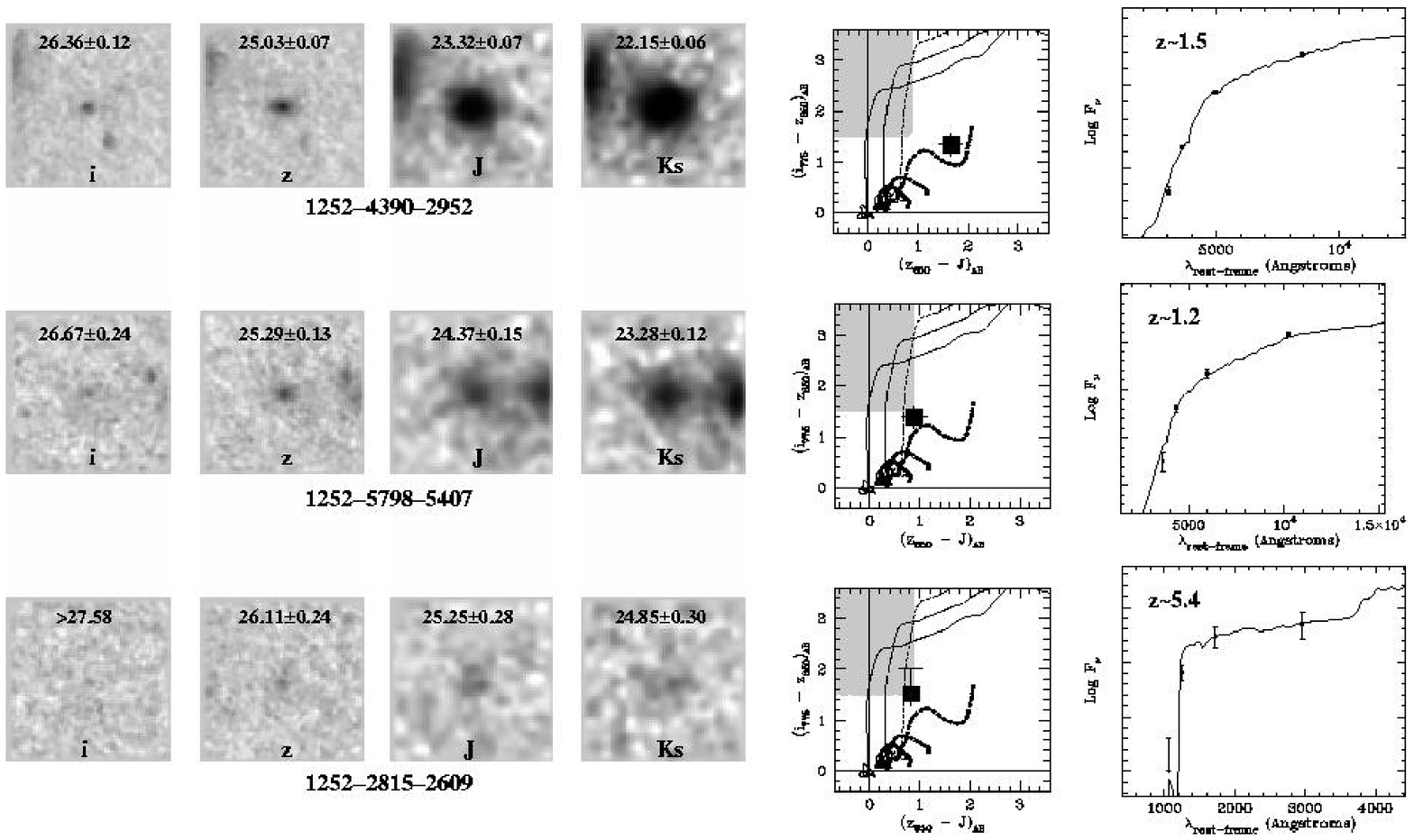}
\caption{$i_{775}z_{850}JK$ postage stamps of several $i-z>1.3$
objects with moderately red $z-J$ colors.  In all three cases, objects
are clearly visible in the infrared, illustrating the basic value of
our infrared data for distinguishing high-redshift objects from low
redshift interlopers.  The top two objects are likely to be
low-redshift ($z\sim1.2-1.5$) ellipticals.  The third object, though
likely at high redshift, did not make our $i$-dropout selection cut.
The postage stamps are 3.0\arcs$\times$3.0\arcs$\,\,$in size.}
\end{figure}

\clearpage
\begin{figure}[p]
\epsscale{0.9}
\plotone{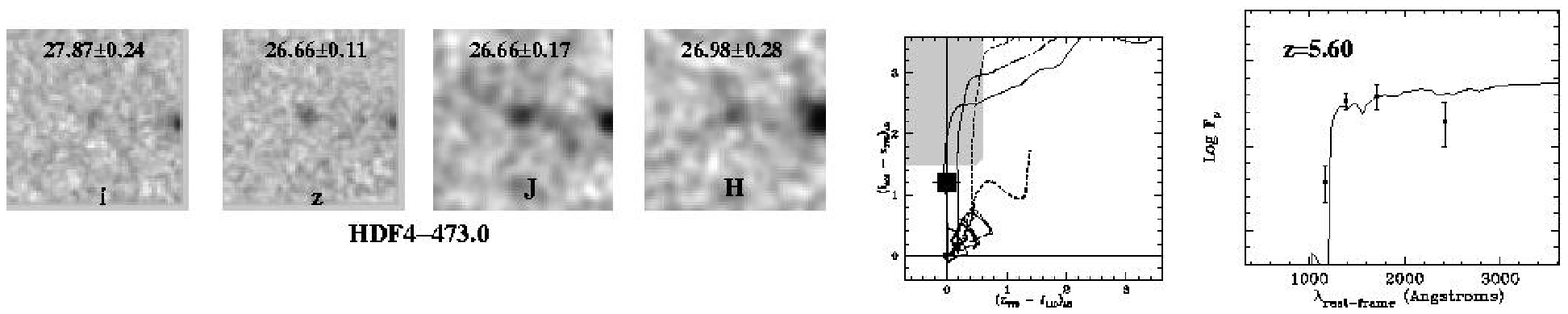}
\caption{$i_{775}z_{850}J_{110}H_{110}$ postage stamps of the $z=5.60$
Weymann et al.\ (1998) object from the HDF North (HDF4-473.0).  While
this object did not meet our selection criteria, it is reassuring to
find a spectroscopically-confirmed object at the low-end of our
$i$-dropout selection range (see \S4) just missing our
$(i_{775}-z_{850})_{AB}>1.5$ cut to the blue.  We measured an
$(i_{775}-z_{850})_{AB}=1.2$ color for this object.}
\end{figure}

\clearpage
\section{Results}

In summary, we find 21 objects over 36 arcmin$^2$ in our RDCS1252-2927
field and 2 objects over 10 arcmin$^2$ in our HDF North field which
satisfy the $i$-dropout criterion ($i-z>1.5$) we defined in the
previous section.  There is deep IR coverage for 12 of these objects
over both fields.  Except for one bright $z_{850,AB}=24.2$ object
which met our $i$-dropout selection cut, $i$-dropouts in our sample
range in magnitude from 25.6 down to our completeness limit
$z_{850,AB}\sim27.3$.  For reference, an average-color (UV power-law
index $\beta=-1.3$) $L_{*}$ object from Steidel et al.\ (1999)'s
$z\sim3$ sample would have $z_{850,AB}\sim26.1$, suggesting a
population of objects with typical luminosities ranging from $\sim$0.3
$L_{*}$ to $\sim$1.5 $L_{*}$.\footnote{Our $z_{850,AB}=24.2$ object
(1252-5224-4599) would therefore be a rather bright and suspiciously
rare $\sim6L_{*}$ object.  Even if this object is a low-redshift
contaminant (and we have no reason to believe that it is), it does not
appear to represent a very large source of contamination, given the
lack of similarly bright objects to $z_{AB}\sim25.6$ and relative
homogeneity of objects faintward of that.} Binning these objects into
0.5 magnitude intervals, we illustrate in Figure 6 how the surface
density of $i$-dropouts varies as a function of magnitude.  Our
typical $i$-dropout has a half-light radius of 0.15\arcs$~$or 0.9 kpc,
though we find them at all sizes ranging from the limit of the PSF
(0.09\arcs) to 0.29\arcs, above which our sample starts to become
incomplete.  We list all objects which lie in our $i$-dropout sample
in Table 1, providing positions, magnitudes, colors, half-light radii,
and the SExtractor stellarity parameter.  Only 13\% ($\sim1-2$
objects) of the $(i_{775}-z_{850})_{AB}>1.5$ objects without IR
magnitudes are likely to be low-redshift contaminants (\S2.3).

\begin{figure}
\epsscale{0.95}
\plotone{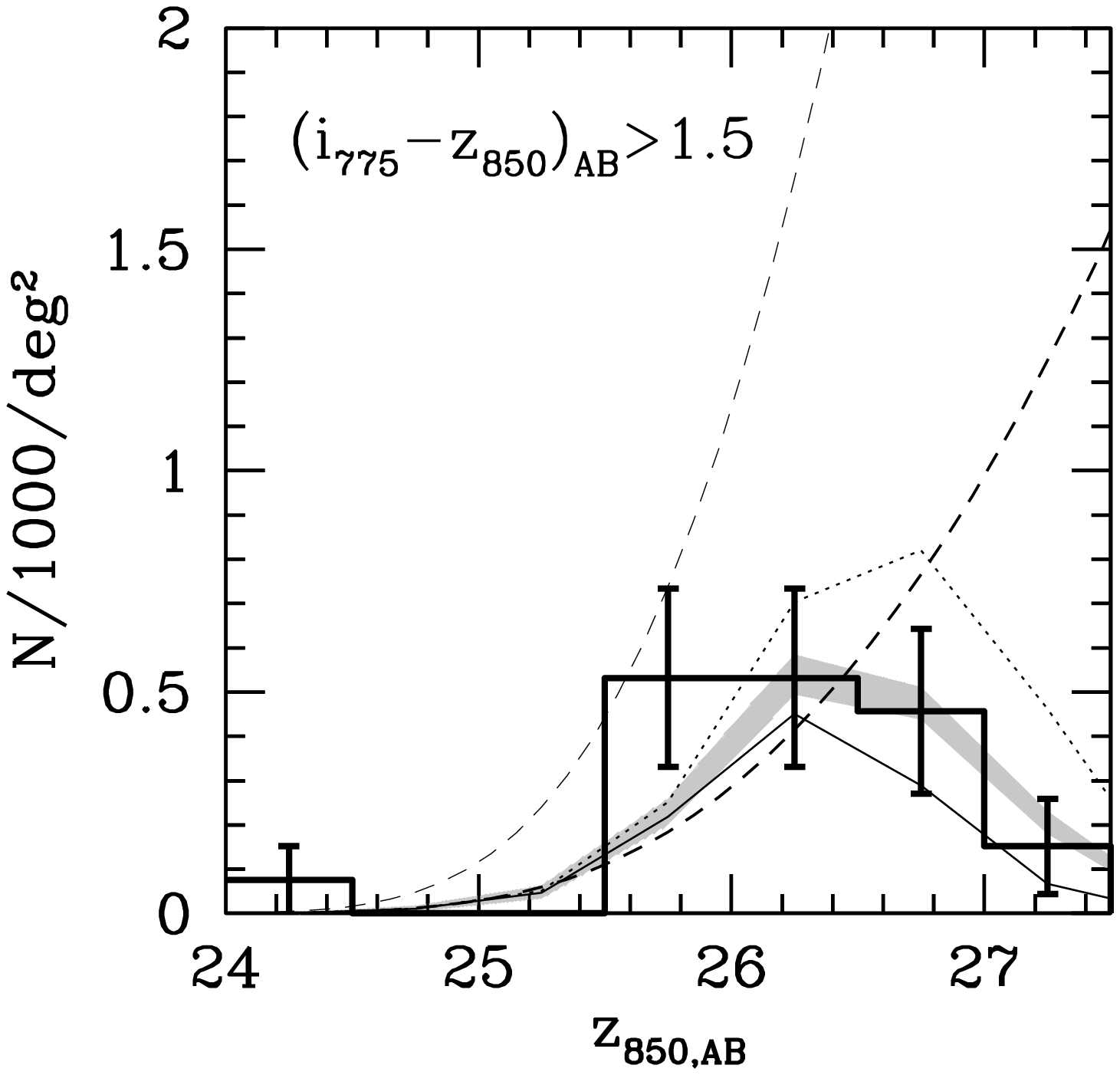}
\caption{Comparison of the number counts for the $i$-dropouts
($i-z>1.5$) observed in our two fields (histogram), corrected by 13\%
for possible contamination, with the no-evolution expectations based
upon our GOODS $z\sim5$ $V$-dropout sample (shaded regions).  We also
include the predictions for two different imaging depths: three orbits
in $i_{775}$/five orbits in $z_{850}$ (solid line) and six orbits in
$i_{775}$/ten orbits in $z_{850}$ (dotted line).  This illustrates how
important the issue of completeness is at these depths.  (Note that
the shaded region above assume that 65\% of our selection area was at
the shallower of these two depths and 35\% of the area was at the
deeper.)  For reference, the thin and thick dashed lines show the
$z\sim3$ Steidel et al.\ (1999) LF placed at $z\sim6$ ($m_{z,*} =
26.1$, $\alpha = -1.6$, $\phi_0 = 0.0025\,\textrm{Mpc}^{-3}$) with no
change in normalization (thin) and a normalization that is 4$\times$
lower (thick), respectively.  (An $i$-dropout selection volume of 800
Mpc$^3$ arcmin$^{-2}$ is assumed for both.)  The observations, while
slightly in excess of the $z\sim5$ no-evolution predictions
(23$\pm$25\%), are consistent with no significant evolution over the
redshift interval $z\sim5-6$.}
\end{figure}

\section{Predictions}

Before getting into a detailed discussion of the luminosity density at
high redshift, perhaps the simplest way to begin interpreting what we
see at $z\sim6$ is to compare it with the $z\sim5$ $V$-dropout sample
we previously selected from the HDF North and South (BBI).  In that
work, we used our cloning machinery (Bouwens, Broadhurst \& Silk
1998a,b; BBI) to project the $z\sim3$ $U$-dropout population to
$z\sim4-5$ for comparison with our $z\sim4$ $B$ and $z\sim5$
$V$-dropout samples.  We found an overall drop in the rest-frame
continuum UV luminosity density (46\% decrease to $z\sim5$) as well as
a modest decrease in the physical size of objects at higher redshifts
relative to the $z\sim3$ population.

We would now like to move this comparison out to higher redshift,
using our HDF $V$-dropout sample to make an estimate for the surface
density of $i$-dropouts on the sky.  Unfortunately, the HDF North
$V$-dropout sample we derived in that work ($(V_{606} - I_{814})_{AB}
> 1.5$, $(V_{606} - I_{814})_{AB} > 3.8, (I_{814} - H_{160})_{AB} -
1.54$, $I_{814,AB} > 24$, and $I_{814,AB} < 27.6$) only contains 1
object brighter than $I_{AB}=26.5$ and hence is not extremely useful
in this regard.  (The fainter $V$-dropouts would not be detectable at
$z\sim6$ to the current depths.)  We therefore resort to use of a much
larger-area $V$-dropout sample from the CDF South GOODS fields (see
Bouwens et al.\ 2003, in preparation).  The virtue of this sample is
both its size (130 objects) and its ACS (PSF FWHM $\sim$0.08\arcs)
resolution.  Our selection criteria for this sample were
$(V_{606}-i_{775})_{AB}>1.7$, $(V_{606}-i_{775})_{AB}>1.1875
(i_{775}-z_{850})_{AB}+1.225$, $(i_{775}-z_{850})_{AB}<1.2$,
$z_{850,AB}<27.2$, where the SExtractor stellarity was less than 0.85
in the $z_{850}$ image (non-stellar with high confidence).  We
selected this sample from essentially the entire area of the CDF
South, i.e., 150 arcmin$^2$ ($\sim$30$\times$ the area of the HDF), so
it should be moderately representative of the universe at $z\sim5$.
Compared to our HDF sample, this $V$-dropout sample has a similar
distribution of redshifts and luminosities.  The methodology for the
generation of this sample remains that of BBI.

With our cloning machinery, we project objects from our
$V_{606}$-dropout sample to higher redshift using the product of the
volume density, $1/V_{max}$ and the cosmological volume.  We
explicitly include pixel-by-pixel k-corrections, cosmic surface
brightness dimming, and PSF variations in calculating the appearance
of these objects to $z\sim5.5$ and beyond.  Assuming no-evolution in
the properties of these samples, we ran Monte-Carlo simulations to
predict the number of $i$-dropouts that would be observed in our
RDCS1252-2927 field and HDF North fields.  We explored searches at two
different depths: three orbits in $i_{775}$, five orbits in $z_{850}$
versus six orbits in $i_{775}$, ten orbits in $z_{850}$.  We then
weighted these predictions by the fractional area observed to these
two different depths (65\% at the shallower depth, 35\% at the
deeper).  The result is that we predict finding 12.1$\pm$1.0 and
4.7$\pm$0.4 $i$-dropouts in our RDCS1252-2927 and HDF North fields,
respectively.  The quoted uncertainties reflect the finite size of the
input $V_{606}$-dropout sample (see Bouwens et al.\ 1998a,b).  Adding
sample variance (simple Poissonian statistics), this works out to
12.1$\pm$3.6 $i$-dropouts in our RDCS1252-2927 field versus the 20
observed (18.7 after the 13\% correction for the low-redshift
contamination) and 4.7$\pm$2.2 $i$-dropouts in our HDF North field
versus the 2 observed (1.9 after correction for contamination),
indicating a slight reduction in the numbers from $z\sim6$ to
$z\sim5$, but more realistically consistent with no evolution (see
Table 2).  We combine the two fields to derive $i$-dropout number
counts and again compare with the no-evolution predictions (Figure 6),
finding 23$\pm$25\% more $i$-dropouts than are predicted using our
no-evolution model.  (Assuming that the luminosity density at $z\sim6$
is proportional to the light in the integrated counts $\int
(10^{-0.4m}) \frac{dN}{dm} dm$, we infer that the luminosity density
is 51$\pm$29\% higher at $z\sim6$ than it is at $z\sim5$.  Removing
the bright $z_{AB}\sim24.2$ object lowers the luminosity increase to
18$\pm$23\%.)  We include on this figure the predictions for the two
depths described above (solid and dotted lines), showing the effect
the assumed depth can have on the predicted numbers.\footnote{We
observe a similarly strong dependence in the data, finding 65\% more
$i$-dropouts arcmin$^{-2}$ in the deeper overlap regions of our
RDCS1252-2927 field than in regions at just half that depth (0.4$^m$
deeper to the same S/N).  This illustrates how important a
consideration incompleteness can be in the magnitude range we are
considering.}  Taking the apparent completeness into account (see also
Figure 8 and the discussion in \S5), we quote an approximate surface
density of $\sim$0.5$\pm$0.2 $i$-dropouts arcmin$^{-2}$ down to
$z_{AB}\sim26.5$.  For reference, we also include a comparison of the
predicted and observed redshift distributions (Figure 7).

\begin{figure}
\epsscale{0.95}
\plotone{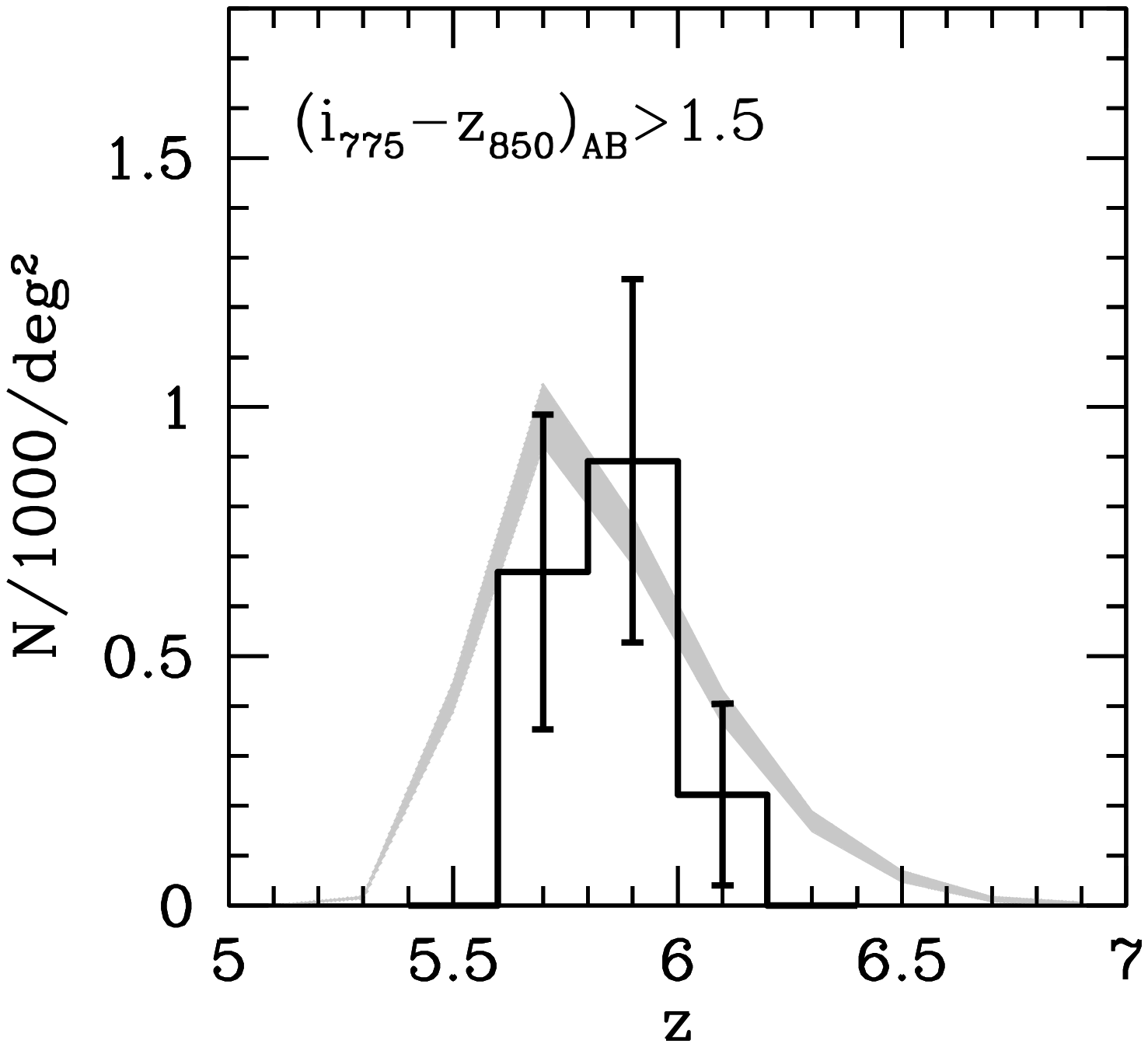}
\caption{Comparison of the estimated redshift distribution for the
observed $i$-dropouts versus that predicted based upon our GOODS
$z\sim5$ $V$-dropout sample (shaded regions).  This shows that the
bulk of the sample is expected to lie between $z\sim5.5$ and
$z\sim6.2$.}
\end{figure}

\section{Estimated Luminosity Density}

In the previous section, we compared the $i$-dropouts we observe with
a no-evolution projection of a wide-area $z\sim5$ $V_{606}$-dropout
sample.  Not only do these simulations give us some gauge of the
evolution across this redshift interval, but they provide an
approximate estimate for the $i$-dropout selection function (assuming
no-evolution in the size, shape, or color distribution).  It is now
relatively straightforward to make three different estimates of the
$z\sim6$ luminosity density.

Our first and most direct estimate comes directly from the comparisons
presented in \S4, where marginal evolution is observed from $z\sim6$
to $z\sim5$.  Linking this with the result from BBI (46\% drop from
$z\sim3$ to $z\sim5$) yields a $20\pm29$\% decrease in the luminosity
density from $z\sim3$ to $z\sim6$ (a 37\% decrease ignoring the bright
$z_{AB}\sim24.2$ object).  This relative decrease can then, in turn,
be expressed as an absolute luminosity density integrating the Steidel
et al.\ (2000) LF down to $z_{AB}\sim27$ or $\sim0.5L_{*}$, where our
$i$-dropouts counts are clearly becoming incomplete.  The result is
$9.8\pm3.6\times10^{25}\emunits$.

Our second estimate closely follows the more standard approach
pioneered by Steidel et al.\ (1999) in deriving the luminosity density
at $z\sim3$.  With this approach, one derives a $UV$-continuum
luminosity function $\phi(M)$ as follows:
\begin{equation}
\phi(M) = \frac{N(m)}{V_{eff} (m)}
\end{equation}
where $M$ is the absolute magnitude at 1600 $\AA$ corresponding to
some $z_{850}$ magnitude $m$ assuming a fixed redshift of 5.9 (the
average redshift for an $i$-dropout, see Figure 7).  The effective
volume then is calculated as a function of magnitude $V_{eff} (m) =
\int_z p(m,z) \frac{dV}{dz} dz$, where $p(m,z)$ is the probability
that an object of magnitude $m$ and redshift $z$ falls in our
$i$-dropout sample and $\frac{dV}{dz}$ is the cosmological volume at
redshift $z$.  The factor $p(m,z)$ contains a whole range of different
selection effects which affect the inclusion of an object in our
sample from the intrinsic distribution of colors to photometric
scatter to the effect of the inherent surface brightnesses on the
completeness of the sample.  Assuming a similar distribution of
surface brightnesses, shapes, and colors to that seen at $z\sim5$, we
can use the Monte-Carlo cloning simulations presented in \S4 to
determine this function.  (Due to the lack of $V_{606}$ dropouts to
probe the selection function $p(m,z)$ at bright magnitudes, we require
that $p(m,z)$ be a strictly decreasing function of magnitude.)  We
present our result in Figure 8.

\begin{figure}
\epsscale{0.95}
\begin{center}
\rotatebox{270}{\resizebox{16cm}{!}{\includegraphics*[47,30][457,400]{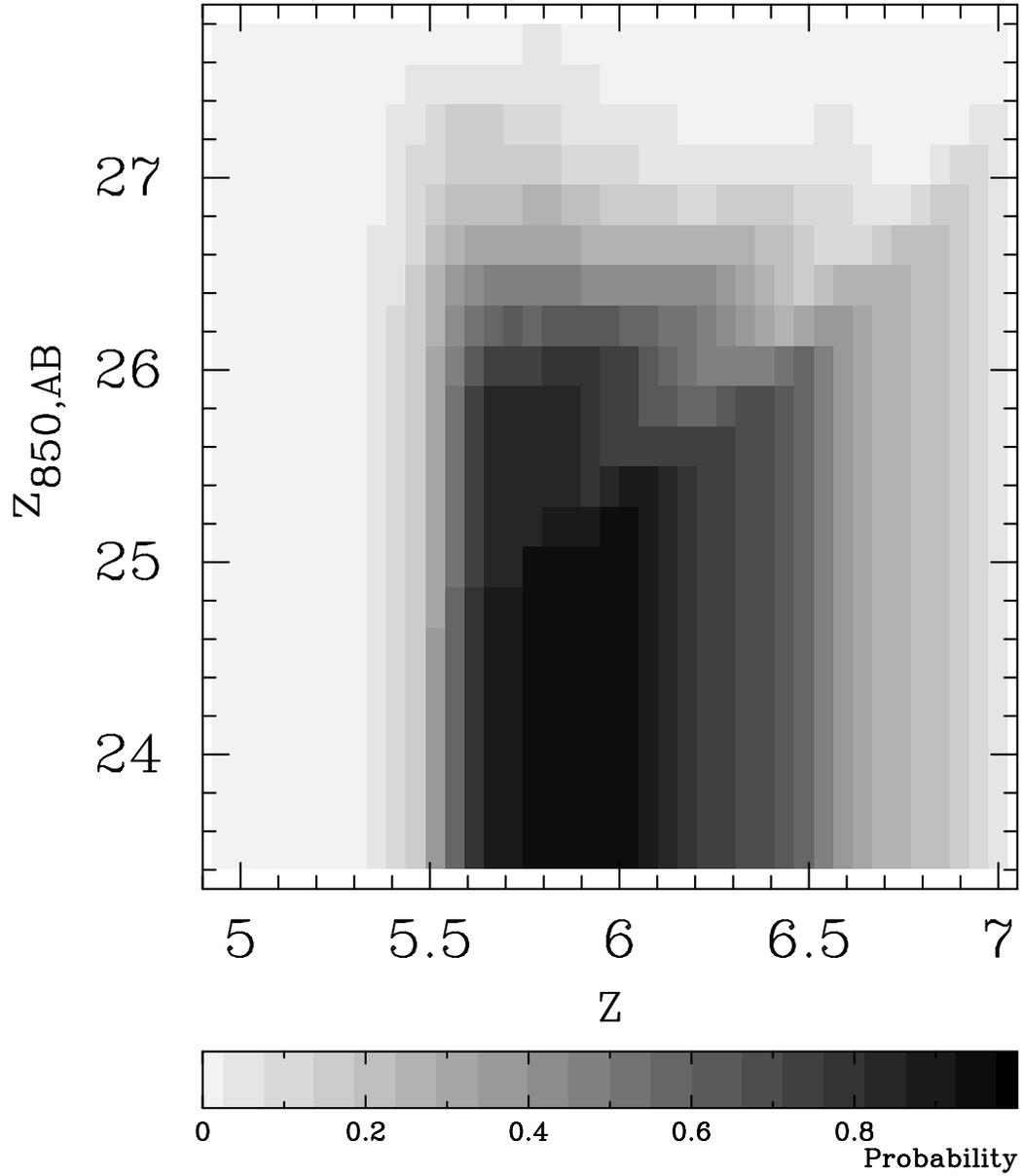}}}
\end{center}
\caption{The probability $p(m,z)$ of some object of $z_{850,AB}$
magnitude $m$ and redshift $z$ being included in our $i$-dropout
sample.  This function was computed from a no-evolution projection of
our wide area $V_{606}$-dropout sample to $z\sim6$ (\S4).}
\end{figure}

For this approach to be effective, the absolute magnitude $M$ has to
be a very tight function of apparent magnitude $m$.  This occurs when
the selection function $p(m,z)$ is a very narrow function of redshift.
Since this is not the case, we rewrite the above expressions as $\int
\phi(M(m,z)) p(m,z) \frac{dV}{dz} dz = N(m)$ where the absolute
magnitude $M$ is a function of the apparent magnitude $m$ and the
redshift $z$.\footnote{In order to convert the apparent magnitudes we
measure to absolute magnitudes, it was necessary for us to take into
consideration possible biases in the measurement of the $z_{850}$
magnitudes.  We can make an estimate for this bias using the
simulations presented in \S4.  Comparing the magnitudes we recover to
those expected based upon an extrapolation of the original $V$-dropout
photometry to $z\sim5.5$, we find an average $\sim$0.2$^m$ faintward
offset in the magnitudes.}  Approximating $\phi(M)$ and $N(m)$ as a
series of step functions, we are able to invert Eq. (3) to solve for
$\phi(M)$ (see Figure 9 for a comparison of the derived LF with the
Steidel et al.\ 1999 $z\sim3$ determination).  Integrating this LF
down to $z_{AB}\sim27$ ($0.5 L_{*}$), we find
$7.2\pm2.5\times10^{25}\emunits$ ($5.9\pm1.8\times10^{25}\emunits$
ignoring the bright $z_{AB}\sim24.2$ object).  This represents a
$39\pm21$\% drop relative to what Steidel et al.\ (1999) report at
$z\sim3$ to a similar limiting luminosity.  Note that for $z\sim6$
$L_{*}$-type objects, the effective survey volume is approximately
$3\times10^4$ Mpc.

\begin{figure}
\epsscale{0.95}
\plotone{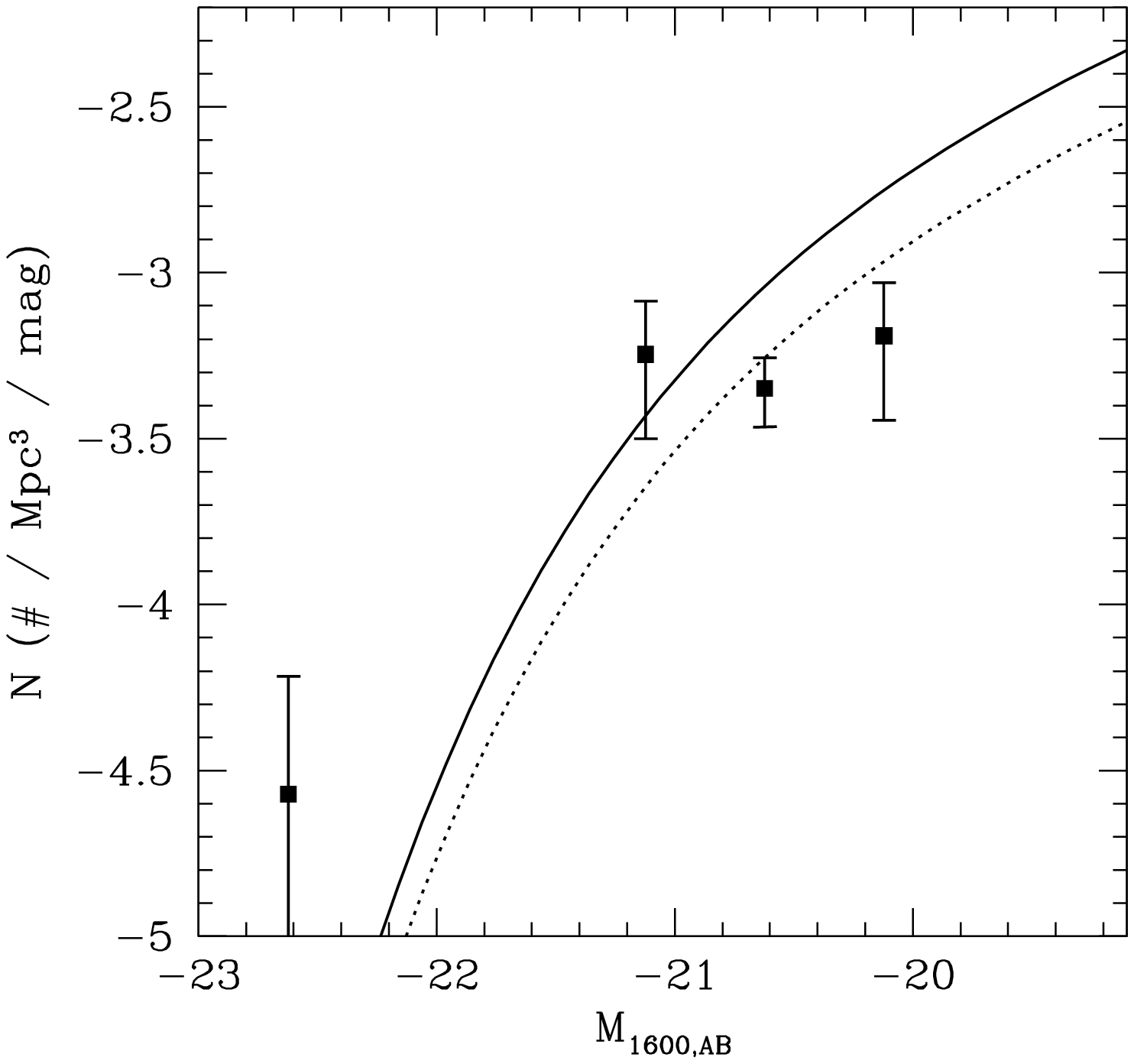}
\caption{The $i$-dropout LF derived from our RDCS1252-2927 and HDF North
fields (solid dots) using a generalized version of Steidel et al.\
(1999)'s $V_{eff}(m)$ technique.  The $z\sim3$ LF of Steidel et al.\
(1999) is superimposed as a solid line ($M_{1600,AB} - M_{1700,AB} =
0.14$ was assumed.)  We also include this same LF with a 39\% lower
normalization (dotted line) to match the observed evolution from
$z\sim6$.  The error bars represent 1$\sigma$ uncertainties.  }
\end{figure}

In the first two approaches presented, there is the implicit
assumption that the selectability of $z\sim6$ objects is similar to
that found at $z\sim5$, both in their colors and their surface
brightnesses, and to a large extent, this is probably true.  The
distribution of $z_{850}-J$ colors (while perhaps a little bluer) are
not that different from the low-redshift expectations (Steidel et al.\
1999; BBI) and similarly for the distribution of surface brightnesses
(as similarities between the predicted and observed number counts in
Figure 6 effectively illustrates).\footnote{Determining both
distributions observationally would require the sort of deep optical
and infrared images promised by the HST Ultra Deep Field.  Lacking
such, one is forced to assume a certain similarity to lower redshift
dropout populations (as we have done here).}  This being said, given
the expectation that higher redshift objects are denser and therefore
of higher surface brightness, it is useful to consider a third
approach where no completeness correction is made.

This enables us to put a lower bound on the $z\sim6$ space density in
the event that these changes have a substantial effect on the
completeness of the $i$-dropout population.  As with the previous two
approaches, we make use of the simulations from \S4.  The difference
is that we only consider objects which actually make it into our
$i$-dropout sample when computing $V_{eff}(m)$, not every object from
our simulations.  The effective volume, $V_{eff}(m)$, is hence
computed as an ensemble average over all inputs observed at magnitude
$m$ and selected as $i$-dropouts.  Putting in our observed surface
densities for these same magnitude intervals (Figure 6), we derive a
UV-continuum luminosity density of $2.7\pm0.6\times10^{25}\emunits$.
We emphasize that relative to our previous estimates, this final
estimate represents a strict lower limit on the luminosity density.

We converted these UV luminosity densities to star formation rate
densities using the relation
\begin{equation}
L_{UV} = \textrm{const}\,\, \textrm{x}\,\, \frac{\textrm{SFR}}{M_{\odot} \textrm{yr}^{-1}} \textrm{ergs}\, \textrm{s}^{-1}\, \textrm{Hz}^{-1}
\end{equation}
where const = $8.0 \times 10^{27}$ at 1500 $\AA$ for a Salpeter IMF
(Madau et al.\ 1998).  The result is $0.0123\pm0.0045\,\sfrd$,
$0.0090\pm0.0031\,\sfrd$, and $0.0034\pm0.0008\,\sfrd$, respectively,
for the three approaches just presented.  Assuming a Schechter
luminosity function with faint end slope $\alpha=-1.6$ and
extrapolating this to the faint end limit yields a SFR density which
is $\sim4\times$ larger.  This works out to an integrated star
formation rate of $\sim0.049\pm0.018\,\sfrd$,
$\sim0.036\pm0.012\,\sfrd$, and $\sim0.014\pm0.003\,\sfrd$,
respectively, for these three approaches.  To put these estimates in
context, we make a comparison with several previous determinations
(Steidel et al.\ 1999; Madau et al.\ 1998; Thompson et al.\ 2001;
Lilly et al.\ 1996; Stanway et al.\ 2003; BBI) in Figure 10,
truncating the observationally-derived LFs at similar faint-end
luminosities.

\begin{figure}
\epsscale{0.95}
\plotone{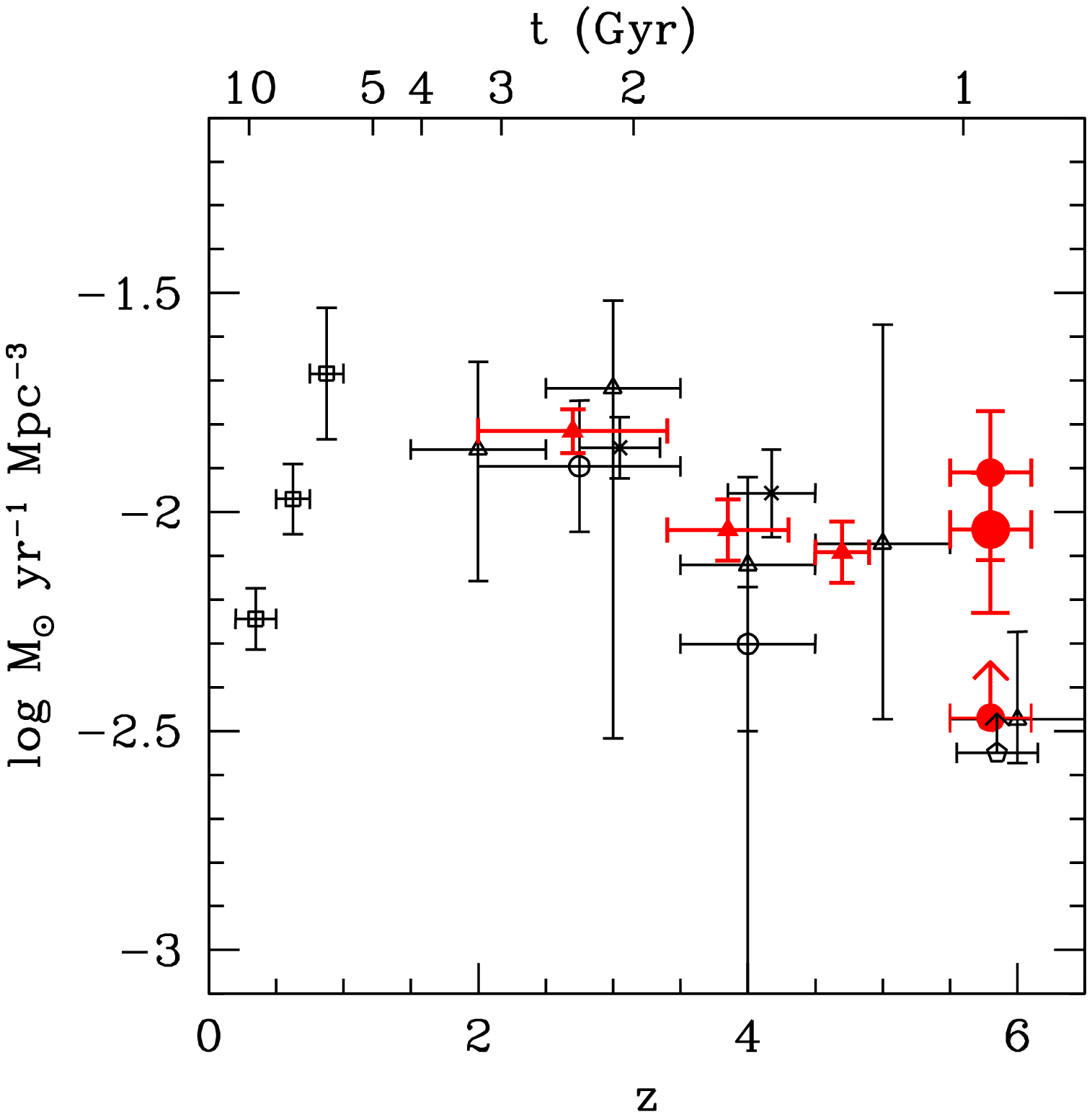}
\caption{A history of the star formation rate density assuming no
extinction correction, integrated down to $0.5L_{*}$.  We include
three determinations from this work (the large solid red circles), the
lower point assuming no incompleteness correction and providing a
reliable lower limit, the slightly larger middle point based on a
generalization of Steidel et al.\ (1999)'s $V_{eff}(m)$ formalism, and
the upper point based on the differential evolution measured from
$z\sim6$ to $z\sim5$ (and linked to $z\sim3$ using the results quoted
in BBI) (see \S5 for details).  The middle point gives our preferred
estimate.  A Salpeter (1955) IMF is used to convert the luminosity
density into a star formation rate (see, for example, Madau et al.\
1998).  Our different estimates provide a nice illustration of the
real uncertainties in the star formation rate density at $z\sim6$,
arising from strong redshift-dependent selection effects.  This topic
will be explored much more extensively in Bouwens et al.\ (2003).
Comparison is made with the previous high redshift determinations of
Lilly et al.\ (1996) (open squares), Madau et al.\ (1998) (open
circles), Steidel et al.\ (1999) (crosses), Thompson et al.\ (2001)
(open triangles), Stanway et al.\ (2003) (open pentagons), and BBI
(solid red triangles).  The top horizontal axis provides the
corresponding age of the universe.  Note the small $\Delta t$ from
$z\sim6$ to $z\sim5$.}
\end{figure}

\section{Discussion}

In this work, the luminosity density at $z\sim6$ is evaluated using
three different procedures.  The first is strictly differential in
nature.  $V_{606}$-dropouts from the wide-area GOODS survey are
projected to $z\sim6$ for comparison with the observed $i$-dropouts.
Adopting previous results to $z\sim5$ (BBI; Thompson et al.\ 2001)
then implies a $\sim$40\% drop in the luminosity density from $z\sim3$
to $z\sim6$.  The second and third procedures, by contrast, are more
direct, relying on a derived selection function to convert the
observed $i$-dropouts into a luminosity density at $z\sim6$.  Relative
to the luminosity density reported at $z\sim3$ by Steidel et al.\
(1999), the resulting drop is a factor of $\sim2$ and $\sim5$,
depending upon the two different assumptions these two procedures make
about the completeness levels (and therefore the presence of low
surface brightness objects at high redshift).  Similarities between
our first two estimates of the luminosity density point to a general
consistency between those methodologies.

Our three estimates have different strengths and weaknesses.  Our
first estimate, for example, depends on the relatively small area BBI
result (HDF-N and HDF-S) and therefore could be quite sensitive to
cosmic variance.  Our second estimate, by contrast, relies upon the
much larger area $z\sim5$ $V$-dropout population (CDFS) to derive the
$z\sim6$ selection function.  Unfortunately, this same $z\sim5$ sample
could be subject to similar selection effects and therefore missing
light, possibly biasing the numbers low.  Our first two estimates
provide accurate estimates of the $z\sim6$ luminosity density in lieu
of large amounts of size or color evolution.  Our final estimate, on
the other hand, is probably much too low except in cases where there
has been dramatic amounts of evolution in the sizes of galaxies.  It
therefore serves as a useful lower limit.  On balance, we prefer our
estimate based on a generalized version of the Steidel et al.\ (1999)
$V_{eff}(m)$ formalism, given the effect of cosmic variance on our
first estimate and the extreme assumptions present in the third (e.g.,
given the small increase in the universal scale factor from $z\sim6$
to $z\sim5$, the change in surface brightnesses should be at most
modest).  We will use it throughout the remainder of the
discussion.\footnote{Ideally speaking, we would estimate the $z\sim6$
luminosity density using a completely differential procedure, scaling
the sizes and colors to match the evolution observed.  Such an
analysis is being performed in a future paper on the dropouts in the
GOODS fields where the samples are substantially larger (Bouwens et
al.\ 2003).}

Before getting into a comparison against previous work, we should take
a more detailed look at the validity of the sample itself,
particularly, with regard to our use of a cluster field.  Two concerns
immediately arise.  The first regards the effect of lensing by the
cluster on $z\sim6$ sources.  We can make a simple estimate of this
effect using the Broadhurst, Taylor, \& Peacock (1995) equation for
the magnification bias, $\bar{M}^{2.5\mu-1}$, where $\mu$ is the slope
of the number counts $\frac{d\log N(m)}{dm}$.  For a faint end slope
$m=0.24$ (a simple consequence of a $z\sim6$ LF where $\alpha=-1.6$)
and an average magnification of $\sim$1.03, this works out to an
expected 1\% \textit{drop} in the surface density of $i$-dropouts
relative to the unlensed case.  The second concern involves the
possibility that $z\sim1.2$ cluster ellipticals might get mixed up in
our $i$-dropout sample.  After all, we decided to observe
RDCS1252-2927 with the $i_{775}$ and $z_{850}$ filters
\textit{because} they straddled the 4000 $\AA$ break.  Fortunately,
the mean $i-z$ color for bright ellipticals in this cluster is
$\sim0.9$ (one can see the cluster as a small overdensity in Figure~1
at $i-z$=0.9 and $z-J$=1.1) and this colour decreases quite markedly
towards faint luminosities (Postman et al.\ 2002; Blakeslee et al.\
2003b).  There is no indication that lower luminosity $z\sim1.2$
cluster members are being scattered into the $i$-dropout sample.

It is interesting to compare the present results with other early work
on the surface density of $i$-dropouts.  Stanway et al.\ (2003), for
example, report finding 8 objects over 146 arcmin$^2$, e.g., 0.05
$i$-dropouts arcmin$^{-2}$ to $z_{AB}\sim25.6$.  We only find one
object ($\approx0.02~i$-dropouts arcmin$^{-2}$) to that same depth, a
result which may not be surprising given the numbers and relative
areas.  In a deeper but smaller area search, Yan et al.\ (2003) report
finding 2.3 objects arcmin$^{-2}$ (27 objects) to $z_{AB}\sim28.0$,
subtracting at most 4 objects from this estimate due to contamination
by stars.  Since the median magnitude of their sources is
$z_{850,AB}\sim27.4$ and their brightest reported source has
$z_{850,AB}\sim26.8$, this works out to $\sim1.2$ objects
arcmin$^{-2}$ to our quoted magnitude limit, somewhat larger than the
0.5$\pm$0.1 objects arcmin$^{-2}$ we find.  We are somewhat uncertain
about the reliability of their identifying $i$-dropouts to
$z_{850,AB}\sim28$ (7 $\sigma$) in a 5-orbit (9540 s) $z_{850}$-band
exposure, given the difficulty we had in identifying $i$-dropouts to a
much brighter limit of $z_{AB}\sim27.3$ ($6 \sigma$) in the deeper
($\sim$10-20 orbit) overlapping regions.  For typical $i$-dropout
sizes ($\sim0.15''$), our analysis would tend to suggest a S/N closer
to 3 at their stated magnitude limit.

Compared to Stanway et al.\ (1995), who estimate a 25\% contamination
fraction due to ellipticals or EROs in an $(i-z)>1.5$ sample, we
estimate a smaller contamination fraction, 13\%, based upon the
measured $z-J$ colors for our sample.  Though hardly significant given
the number of objects involved, such a difference clearly follows the
trend toward lower early-type fractions at fainter magnitudes.  We
noted similarly small contamination fractions in our analysis of the
$V$-dropouts from the HDF North (BBI).

Down to our magnitude limit $z_{AB}\sim27.3$, we find a star formation
rate density at $z\sim6$ of $0.0090\pm0.0031\sfrd$, $\sim$2$\times$
lower than similar estimates at $z\sim3$ from Steidel et al.\ (2003)
and Madau et al.\ (1998).  This estimate is nearly $14\times$ Stanway
et al.\ (2003)'s quoted result ($6.7\pm2.7\times10^{-4}\sfrd$).  While
a portion of this difference can be attributed to the different depths
of our searches (the present census going $\sim$1.4 magnitudes further
down the luminosity function), there are significant differences in
the effective volumes we assume, the present study correcting for the
significant incompleteness at $z\sim6$ resulting from surface
brightness dimming while Stanway et al.\ (2003) do not make such a
correction.  For both this reason and our greater overall depth, the
present study represents a clear improvement over the Stanway et al.\
(2003) estimate.  From $z\sim5$ to $z\sim6$, we find minimal evidence
for dramatic evolution, consistent with the small amount of cosmic
time available across this interval ($\sim$0.2 Gyr from $z\sim6$ to
$z\sim5$ and $\sim$1 Gyr from $z\sim6$ to $z\sim3$).  In general, the
trends we find are consistent with the gradual decline in star
formation density previously reported from $z\sim3$ to $z\sim5$ (Madau
et al.\ 1996; Steidel et al.\ 1999; Thompson et al.\ 2001; Lehnert et
al.\ 2003; Deltorn et al.\ 2003; BBI) or as found in recent work on
Lyman-alpha emitters, where a deficit is claimed at $z\sim5.7$
relative to the $z\sim3$ population (Maier et al.\ 2003).

Since recent work from absorption line studies on $z>5.8$ QSOs
indicates that the universe may have been reionized near a redshift of
$z\sim6$ (Fan et al.\ 2001; Becker et al.\ 2001; Fan et al.\ 2002), it
is relevant to compare the star formation rate we determine here with
that needed to reionize the universe at $z\sim6$.  For the latter, we
use the following relation from Madau, Haardt, \& Rees (1999),
correcting it for the baryon density ($\Omega_b h^2 = 0.0224$) derived
from the recent WMAP results (Bennett et al.\ 2003) and shifting it to
$z\sim6$:
\begin{equation}
\dot{\rho_{*}} \approx (0.052 \,\sfrd)\,\left(\frac{0.5}{f_{\rm
esc}}\right)\,C_{30} \left(\frac{1+z}{7}\right)^{3}.
\end{equation}
where $\dot{\rho_{*}}$ is the star formation rate density, $C_{30}$ is
the $\HI$ concentration factor $(1/30) \left\langle \rho_{\HI}^{2}
\right\rangle$ $\left\langle \rho_{\HI}\right\rangle^{-2}$, and
$f_{\rm esc}$ is the fraction of ionizing radiation escaping into the
intergalactic medium.  Given the large observational and theoretical
uncertainties in the exact values of $C_{30}$ and $f_{\rm esc}$, the
star formation rate densities inferred here at $z\sim6$
($\sim$0.0090$\sfrd$ observed, $\sim$0.036$\sfrd$ extrapolating the LF
to the faint limit) are well within the range of that needed to
reionize the universe at $z\sim6$ ($\sim$30\% lower than the fiducial
value given in Eq. (3)).  Whether or not the objects we observe at
$z\sim6$ are sufficient to reionize the universe or this ionizing
radiation is provided by a completely different population of objects
(e.g., Madau 1998), it seems clear that this ionizing radiation does
not come from $z\sim5-6$ AGNs (Haiman, Madau, \& Loeb 1998).

\section{Summary}

We use deep ACS WFC $i$+$z$ observations of RDCS1252-2927 and the HDF North
to search for $z\sim6$ candidates by looking for a strong Lyman break
across the $i+z$ passbands in fields with deep IR imaging.  We augment
this with deep infrared imaging to derive $z-J$ colors for the
$z\sim6$ candidates to help distinguish them from lower redshift
interlopers where possible.  We compare our findings with a
no-evolution projection of a wide-area $V_{606}$-dropout sample to
$z\sim6$ and then use these simulations to make three different
estimates of the rest-frame continuum $UV$-luminosity density at
$z\sim6$.
\begin{itemize}
\item{To $z_{AB}\sim27.3$, we find 21 $i$-dropouts ($i-z>1.5$;
$6\sigma$ detections) over 36 arcmin$^2$ in our RDCS1252-2927 field
(one object is lensed) and 2 $i$-dropouts over 10 arcmin$^2$ in our
HDF North field.  This is equivalent to $\sim0.5\pm0.1$
object arcmin$^{-2}$ down to our magnitude limit $z_{AB}\sim27.3$, or
$\sim0.5\pm0.2$ object arcmin$^{-2}$ down to $z_{AB}\sim26.5$ corrected
for completeness.}
\item{Compared to an average-color (UV power-law index $\beta=-1.3$)
$U$-dropout in the Steidel et al.\ (1999) $z\sim3$ sample, the
$i$-dropouts we find range in luminosity from $\sim$0.3 $L_{*}$ to
$\sim$1.5 $L_{*}$ (with the exception of one very bright
$z_{850,AB}=24.2$ ($6L_{*}$) object which also meets our selection
criterion).  Our typical $i$-dropout has a half-light radius ranging
from 0.09\arcs$~$to 0.29\arcs, or 0.5 kpc to 1.7 kpc, for a $\Omega_M
= 0.3$, $\Omega_{\Lambda}=0.7$, $h = 0.7$ cosmology.}
\item{Using our deep infrared data to constrain the spectral slopes
redward of the break, we find that all 12 of our $i$-dropout
candidates ($i-z>1.5$), for which we have IR data, have blue ($<0.8$)
$z-J$ colors.  Using the Kaplan-Maier estimator on all extended
$i-z>1.3$ sources, we estimate that only 13\% of such $i-z>1.5$
sources are low-redshift interlopers.  Assuming spectroscopic
confirmation of the two-color dropout technique to $z\sim6$ (e.g., Fan
et al.\ 2001; Bunker et al.\ 2003), this demonstrates that an $i-z$
selection can be an effective way of uncovering a population of high
redshift objects (see also Stanway et al.\ 2003).}
\item{Over the 21 arcmin$^2$ where we had infrared coverage, we
identified only one red ($i-z>1.3$) point-like object with $z-J$
colors consistent with being a $z\sim5.5-6$ AGN.  Since in total we
identify 7 point-like objects with $z_{AB}>25$ and
$(i_{775}-z_{850})_{AB}>1.3$ and only one out of four objects (25\%)
with IR coverage has $z-J$ colors less than 0.8, this works out to an
estimated surface density of $\sim$0.04$\pm$0.03 $z\sim5.5-6.3$ AGNs 
arcmin$^{-2}$ down to our magnitude limit of $z_{AB}\sim27.3$.}
\item{Comparing the number of $i$-dropouts with a no-evolution
projection of our $z\sim5$ $V_{606}$-dropout sample from CDF South
GOODS, we estimate the evolution in rest-frame continuum UV luminosity
density from $z\sim6$ to $z\sim5$.  We find 23$\pm$25\% more
$i$-dropouts ($\sim$21) than are predicted (17) once a correction for
the contamination rate is made, consistent with no strong evolution
over this redshift interval.  Redoing this increase in terms of
integrated luminosity, this increase amounts to 51$\pm$29\% (or
$18\pm23$\% removing the one very bright $z_{AB}\sim24.2$ object).
Adopting previous results to $z\sim5$ (BBI; Thompson et al.\ 2001),
this works out to a mere 20$\pm$29\% drop in the luminosity density
from $z\sim3$ to $z\sim6$.}
\item{Using simulations based on a set of CDF South
$V_{606}$-dropouts, we estimate the selection function $p(m,z)$ for
our $i$-dropout sample.  Then, via a generalized version of Steidel et
al.\ (1999)'s $V_{eff}(m)$ formalism, we calculate the $UV$-luminosity
function for this sample, and integrate it down to $z_{AB}\sim27.0$
($\sim0.5L_{*}$).  The rest-frame $UV$-luminosity density we derive
($7.2\pm2.5\times10^{25}\emunits$) is $39\pm21$\% lower than
Steidel et al.\ (1999) found to a similar limiting luminosity,
consistent with the above estimate.  This is our preferred estimate.}
\item{The previous two approaches assume no large change in the
selectability (color or surface brightness distribution) of dropouts
from $z\sim5$ to $z\sim6$.  Given the expectations (observational and
theoretical) that dropouts may have higher surface brightnesses at
$z\sim6$ than $z\sim5$, it is useful to make a third estimate for the
luminosity density from the simulations, but this time assuming no
incompleteness.  Running through the numbers, we find
$2.7\pm0.6\times10^{25}\emunits$ for the \textit{observed}
$i$-dropouts, $\sim$5$\times$ lower than the value reported by Steidel
et al.\ (1999) to a similar luminosity.  This third approach provides
a reliable lower limit to the luminosity density.}
\item{Converting the luminosities densities we infer into star
formation rate densities using standard assumptions (e.g., Madau et
al.\ 1998), we find an integrated star formation rate density of
$0.0123\pm0.0045\,\sfrd$, $0.0090\pm0.0031\,\sfrd$, and
$0.0034\pm0.0008\,\sfrd$, respectively, for the three approaches just
presented (or $\sim0.049\pm0.018\,\sfrd$, $\sim0.036\pm0.012\,\sfrd$,
and $\sim0.014\pm0.003\,\sfrd$ extrapolating the observations to low
luminosities using a Schechter function with faint-end slope
$\alpha=-1.6$).}
\item{Our preferred estimate for the rest-frame continuum UV
luminosity density and star formation rate density at $z\sim6$ are
$7.2\pm2.5\times10^{25}\emunits$ and $0.0090\pm0.0031\,\sfrd$
($\sim0.036\pm0.012\,\sfrd$ extrapolating the luminosity function to
the faint limit), respectively.  This represents a 39$\pm$21\% drop
from $z\sim3$ to $z\sim6$.}
\item{The $z\sim6$ rest-frame continuum $UV$-luminosity densities we
infer are well within the expected range needed for reionization, for
canonical assumptions about the $\HI$ clumping factor and the fraction
of UV radiation escaping into the intergalactic medium.}
\end{itemize}

The combination of deep ACS $i+z$ and ground-based IR imaging have
been shown to be a very effective means of isolating high redshift
objects and studying their properties.  We will be following up this
analysis with an investigation of the shallower, but larger area, data
in the CDF South and HDF North from the GOODS program (Bouwens et al.\
2003).  \acknowledgements

We extend a special thanks to Mark Dickinson for providing us with his
fully reduced NICMOS images of the HDF North.  We would also like to
acknowledge T. Allen, K. Anderson, S. Barkhouser, S. Busching,
A. Framarini, and W.J. McCann for their invaluable contributions to
the ACS Investigation Definition Team (IDT).  This research has made
use of the NASA/IPAC Extragalactic Database (NED) which is operated by
the Jet Propulsion Laboratory, California Institute of Technology,
under contract with the National Aeronautics and Space Administration.
ACS was developed under NASA contract NAS 5-32865.  RJB, GDI, and the
ACS IDT acknowledge the support of NASA grant NAG5-7697.

{}

\appendix

\begin{center}
\large
\textbf{Appendix}
\normalsize
\end{center}
\section{Noise}

The drizzling and resampling procedure we employ (and generally
employed with HST data sets) to produce our fully coaligned optical
and infrared data set naturally introduces a certain correlation into
the noise of this data set.  To estimate the true amplitude of the
noise--distinct from the single pixel RMS--as well as the effective
noise kernel, RMS fluctuations within apertures of increasing sizes
are measured, after masking out $>5\sigma$ detections, up to a scale
where there is no obvious additional power.  A noise model (amplitude
and kernel) is then constructed which provides a plausible fit to the
observations.  This model is used throughout to estimate errors in the
photometry.

\section{$z\sim6$ AGN}

We found one $(i_{775}-z_{850})_{AB}\sim1.3$, $(z_{850}-J)_{AB}<-0.1$
stellar object consistent with a $z\sim5.5-6$ AGN-identification over
the area where we had infrared coverage ($\sim$21 arcmin$^2$) (Figure
11).  Since in total we identify 7 point-like objects with $z_{AB}>25$
and $(i_{775}-z_{850})_{AB}>1.3$ and only one out of the four objects
with IR coverage (25\%) had $(z_{850}-J)_{AB}$ colors less than 0.8,
this works out to an estimated surface density of $\sim$0.04$\pm$0.03
objects arcmin$^{-2}$ down to our magnitude limit of
$z_{850,AB}\sim27.3$.

\begin{figure}[p]
\epsscale{0.95}
\plotone{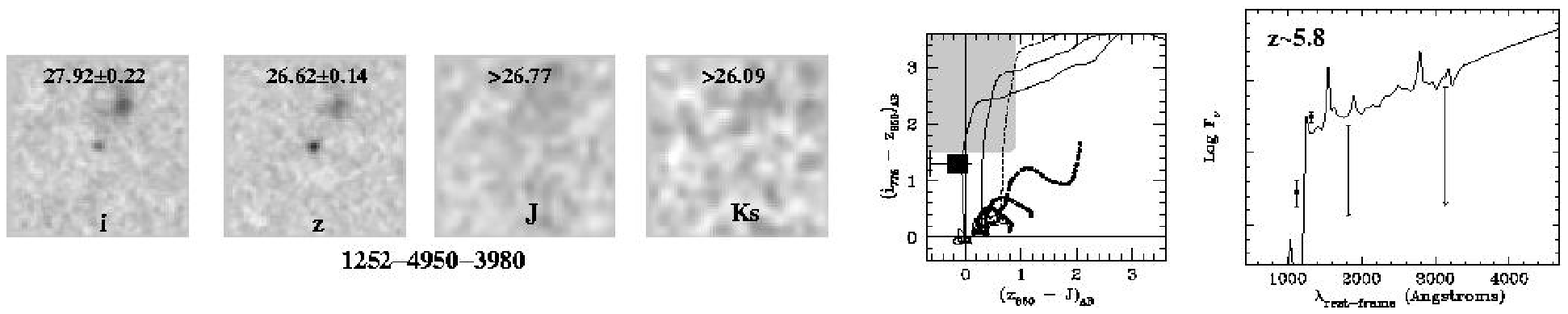}
\caption{$i_{775}z_{850}JK$ images of a $z\sim6$ AGN candidate found
in our RDCS1252-2927 field.  The above object is the only such candidate over
the 21 arcmin$^2$ where we have both optical and infrared imaging.
The position in $i-z$, $z-J$ color-color space is included along with
a plausible fit to an SED.}
\end{figure}

\newpage

\begin{deluxetable}{lrrrrrrrrr}
\tablewidth{0pt}
\tabletypesize{\footnotesize}
\tablecaption{$z\sim6$ Sample.  Objects from RDCS1252-2927 and the HDF North
which satisfy our $(i_{775}-z_{850})_{AB}>1.5$ $i$-dropout
criterion.\label{tbl-1}}
\tablehead{
\colhead{Object ID} &
\colhead{Right Ascension} & \colhead{Declination} &
\colhead{$z_{850}$\tablenotemark{a}} & \colhead{$i - z$\tablenotemark{b}}
& \colhead{$z-J$\tablenotemark{c}} & \colhead{$V-z$} 
& \colhead{$I-z$} & \colhead{S/G\tablenotemark{d}} & \colhead{$r_{hl}$(\arcs)}}
\startdata
1252-5224-4599\tablenotemark{e} & 12:52:56.8880 & -29:25:55.503 & 24.2$\pm$0.1 & 1.5 & 0.1 & -- & -- & 0 & 0.29\\
1252-2134-1498 & 12:52:45.3816 & -29:28:27.105 & 25.6$\pm$0.2 & $>$2.1 & 0.4 & -- & -- & 0.01 & 0.18\\
1252-2585-3351 & 12:52:52.2832 & -29:28:04.743 & 25.7$\pm$0.1 & 2.0 & 0.1 & -- & -- & 0 & 0.20\\
1252-6031-966 & 12:52:43.3793 & -29:25:17.474 & 25.7$\pm$0.1 & 2.0 & -- & -- & -- & 0.30 & 0.11\\
1252-562-2836 & 12:52:50.2971 & -29:29:43.226 & 25.8$\pm$0.1 & 2.0 & -- & -- & -- & 0.06 & 0.14\\
1252-6798-302 & 12:52:40.8435 & -29:24:39.033 & 25.8$\pm$0.2 & 1.7 & -- & -- & -- & 0.04 & 0.16\\
1252-4720-6466 & 12:53:03.8342 & -29:26:20.835 & 25.8$\pm$0.1 & 1.5 & -- & -- & -- & 0.02 & 0.12\\
1252-5058-5920 & 12:53:01.7448 & -29:26:03.873 & 25.9$\pm$0.2 & $>$1.8 & -0.4 & -- & -- & 0 & 0.19\\
1252-277-3385 & 12:52:52.3987 & -29:29:57.525 & 26.0$\pm$0.2 & $>$1.6 & -- & -- & -- & 0 & 0.19\\
1252-5377-2621 & 12:52:49.5122 & -29:25:47.721 & 26.0$\pm$0.2 & $>$1.6 & $<$-0.1 & -- & -- & 0 & 0.22\\
1252-7065-6877 & 12:53:05.4237 & -29:24:26.222 & 26.1$\pm$0.2 & 1.6 & -- & -- & -- & 0.01 & 0.14\\
1252-3544-3542\tablenotemark{*} & 12:52:53.0237 & -29:27:16.801 & 26.2$\pm$0.2 & 1.6 & 0.1 & -- & -- & 0 & 0.23\\
1252-2439-3364 & 12:52:52.3340 & -29:28:12.030 & 26.4$\pm$0.2 & $>$1.7 & $<$-0.2 & -- & -- & 0 & 0.17\\
1252-7005-1697 & 12:52:46.1876 & -29:24:28.846 & 26.5$\pm$0.2 & $>$1.5 & -- & -- & -- & 0 & 0.13\\
1252-3729-4565 & 12:52:56.7460 & -29:27:07.631 & 26.7$\pm$0.2 & $>$1.8 & 0.4 & -- & -- & 0 & 0.12\\
1252-7313-6944 & 12:53:05.6814 & -29:24:13.836 & 26.7$\pm$0.2 & $>$1.6 & -- & -- & -- & 0.09 & 0.09\\
1252-1199-3650 & 12:52:53.4200 & -29:29:11.443 & 26.7$\pm$0.2 & $>$1.7 & 0.1 & -- & -- & 0 & 0.12\\
1252-5399-4314 & 12:52:55.7957 & -29:25:46.724 & 27.0$\pm$0.2 & $>$1.6 & $<$0.0 & -- & -- & 0 & 0.11\\
1252-3497-809 & 12:52:42.7537 & -29:27:18.894 & 27.0$\pm$0.2 & $>$1.6 & -- & -- & -- & 0 & 0.11\\
1252-4375-1877 & 12:52:46.8536 & -29:26:37.733 & 27.2$\pm$0.2 & 1.6 & $<$0.7 & -- & -- & 0.60 & 0.09\\
1252-3684-528 & 12:52:41.6802 & -29:27:09.550 & 27.3$\pm$0.2 & $>$1.5 & -- & -- & -- & 0.16 & 0.09\\
HDFN-2135-3286 & 12:36:49.9368 & 62:13:55.671 & 26.5$\pm$0.2 & $>$1.9 & $<$-0.6\tablenotemark{**} & $>$2.1 & 1.1 & 0.02 & 0.14\\
HDFN-4965-4355 & 12:36:30.7457 & 62:12:53.371 & 27.0$\pm$0.2 & $>$1.5 & -- & $>$1.3 & -- & 0.01 & 0.10\\
\enddata
\tablenotetext{a}{AB Magnitudes}
\tablenotetext{b}{All limits are 2$\sigma$.}
\tablenotetext{c}{Here the $J$ band alternatively refers to the ISAAC $J$ band and the NICMOS $J_{110}$ filter depending upon which field the object is found.}
\tablenotetext{d}{SExtractor stellarity parameter, for which 0 =
extended object and 1 = point source}
\tablenotetext{e}{The ``1252-'' prefix denotes an object from RDCS1252-2927.}
\tablenotetext{*}{This object appears to be lensed and therefore is excluded from our sample.}
\tablenotetext{**}{This is $(z-J_{110})_{AB}$.}
\end{deluxetable}

\begin{deluxetable}{cccc}
\tablewidth{0pt}
\tablecaption{The number of $i$-dropouts found in our samples versus
no-evolution predictions.  Two different $1\sigma$ uncertainties are
quoted on all predictions, the first based on the finite size of our
$V$-dropout sample (130 objects) (Bouwens et al.\ 1998a,b) and the
second based on sample variance (simple Poissonian errors).
\label{dropouts}} 
\tablehead{\colhead{Data set} & \colhead{Observed} & \colhead{No evolution
Prediction}} 
\startdata 
RDCS1252-2927 & 18.7\tablenotemark{a} & 12.1$\pm$1.0$\pm$3.5 \\
 HDF North & 1.9\tablenotemark{a} &
4.7$\pm$0.4$\pm$2.2 \enddata \tablenotetext{a}{Corrected for the
expected 13\% contamination rate.}
\end{deluxetable}

\end{document}